\documentclass[journal=jacsat,manuscript=article,layout=twocolumn]{achemso}

\usepackage[version=3]{mhchem} 
\usepackage{nicefrac}
\usepackage{amssymb}
\usepackage{color,soul}
\usepackage{newtxtext}
\usepackage[scaled=0.9]{helvet}
\usepackage[fontsize=9pt]{fontsize}
\usepackage{amsmath}
\usepackage{amsfonts}
\usepackage{xcolor}
\usepackage{chemformula}
\usepackage[flushleft]{threeparttable}
\usepackage{multirow}
\usepackage{array}
\usepackage{ascii}
\usepackage{lettrine}
\usepackage{float}
\usepackage{placeins}

\usepackage{achemso}
\usepackage{multicol}
\setkeys{acs}{maxauthors = 10, etalmode = truncate, doi=true}

\renewcommand{\=}{\mbox{\,=\,}}
\newcommand{\im}{\mathrm{Im}}

\author{Nikhil S. Chellam}
\affiliation[Northwestern University]{Department of Chemical \& Biological Engineering, Northwestern University, Evanston, Illinois, 60208, United States}
\alsoaffiliation[IIN]{International Institute for Nanotechnology, Northwestern University, Evanston, Illinois, 60208, United States}
\author{Subhajyoti Chaudhuri}
\affiliation[Northwestern University]{Department of Chemistry, Northwestern University, Evanston, Illinois, 60208, United States}
\alsoaffiliation[IIN]{International Institute for Nanotechnology, Northwestern University, Evanston, Illinois, 60208, United States}
\author{Abhisek Ghosal}
\affiliation[Northwestern University]{Department of Chemistry, Northwestern University, Evanston, Illinois, 60208, United States}
\author{Sajal Kumar Giri}
\affiliation[ISM Dhanbad]{Department of Chemistry and Chemical Biology, Indian Institute of Technology (ISM) Dhanbad, Dhanbad 826004, India}
\email{sajalgiri@iitism.ac.in}
\author{George C. Schatz}
\affiliation[Northwestern University] {Department of Chemical \& Biological Engineering, Northwestern University, Evanston, Illinois, 60208, United States}
\alsoaffiliation[Northwestern Chem]{Department of Chemistry, Northwestern University, Evanston, Illinois, 60208, United States}
\alsoaffiliation[IIN]{International Institute for Nanotechnology, Northwestern University, Evanston, Illinois, 60208, United States}
\email{g-schatz@northwestern.edu}

\title[review title]
  {Density Functional Tight-Binding Enables Tractable Studies of Quantum Plasmonics}

\abbreviations{IR,NMR,UV}
\keywords{American Chemical Society, \LaTeX}

\begin{document}
\abstract{Routine investigations of plasmonic phenomena at the quantum level present a formidable computational challenge due to the large system sizes and ultrafast timescales involved. This Feature Article highlights the use of density functional tight-binding (DFTB), particularly its real-time time-dependent formulation (RT-TDDFTB), as a tractable approach to study plasmonic nanostructures from a purely quantum mechanical purview. We begin by outlining the theoretical framework and limitations of DFTB, emphasizing its efficiency in modeling systems with thousands of atoms over picosecond timescales. Applications of RT-TDDFTB are then explored in the context of optical absorption, nonlinear harmonic generation, and plasmon-mediated photocatalysis. We demonstrate how DFTB can reconcile classical and quantum descriptions of plasmonic behavior, capturing key phenomena such as size-dependent plasmon shifts and plasmon coupling in nanoparticle assemblies. Finally, we showcase DFTB's ability to model hot carrier generation and reaction dynamics in plasmon-driven \ch{H2} dissociation, underscoring its potential to model photocatalytic processes. Collectively, these studies establish DFTB as a powerful, yet computationally efficient tool to probe the emergent physics of materials at the limits of space and time.}

\section{Introduction}
\lettrine{O}{ne of the earliest} reports on the anomalous coloration of small metal particles dates to a 700 B.C. Assyrian tablet describing a recipe for red glass by adding gold.\cite{holmyard1925} Over two millennia later, Michael Faraday attributed this vivid coloration to light scattering in colloidal gold, while Gustav Mie developed the first complete analytical description for such with spherical particles soon after.\cite{Faraday1857, Mie1908} This phenomenon ultimately arises from gold, silver, and many other metal particles' ability to sustain localized surface plasmons at the nanoscale; excitation of these collective oscillations of conduction band electrons produces a strong resonant absorption band in the UV-visible-IR region, giving rise to the brilliant colors found in these colloidal solutions. The plasmon resonance wavelength can be further tuned by modifying the nanoparticle's size, shape, and surrounding dielectric environment\cite{Schatz2003, Halas2011} – a feature widely exploited in diverse fields including catalysis,\cite{Zhou2018, Al-Zubeidi2023, Cortes2020, Kumar2019} optics,\cite{Zheng2021,Khorasaninejad2016,Yu2012} (bio)chemical sensing,\cite{Elghanian1997, Jeanmaire1977, Pockrand1983, Wu2019} and therapeutics.\cite{Hirsch2003, Dam2012, Choo2021}

Today, advances in colloidal synthesis and nanofabrication techniques have enabled the creation of nanoparticles with precisely controlled morphologies and consequent optical properties.\cite{Laramy2019, Kalsin2006, Jin2001, Boles2016} Complementary theoretical studies have revealed detailed correlations between particle characteristics and their optical responses difficult to characterize \emph{via} experimentation alone.\cite{Schatz2003, Santiago2020, Rossi2020} In turn, the interplay between theoretical analysis and experimental discovery has effected breakthroughs in metamaterial design,\cite{Sample2021, Guan2023-2} advanced lasing technologies,\cite{Guan2023, Wang2018, Yang2015} catalysis,\cite{Christopher2011, Mukherjee_2013} and ultrasensitive chemical detection.\cite{Haes2004}

Richard Feynman's 1959 claim that ``there's plenty of room at the bottom" heralded the nanotechnology revolution, yet a fundamental tension has persisted in theoretical quantum chemistry for the nearly 70 years since: what constitutes ``nano" spans orders of magnitude in complexity. Plasmonic phenomena represent a different scale altogether. These collective excitations emerge only in systems containing hundreds to thousands (to even millions) of atoms -- massive by quantum mechanical standards -- and their picosecond-scale dynamics create a computational gauntlet that even our most advanced hardware struggles to navigate. While time-dependent Kohn-Sham density functional theory (TDDFT) stands as the nominal gold standard for analyzing photochemical processes, it hits a wall with particles larger than 1-2 nm and times beyond tens of femtoseconds.\cite{Wu2020, Mortensen2014, Herring2023} This scheme becomes intractable for particle sizes on the order of hundreds to thousands of atoms, becoming \textit{de facto} unusable for many experimentally-observable plasmonic systems. Methods that bring quantum mechanical calculations to longer spans of space and time would thus be valuable to describe the length scales for describing plasmons and the timescales governing their emergent chemistry.

First conceptualized in 1986\cite{Seifert1986} with extensions implemented soon after,\cite{Foulkes_1989, Porezag1995, Elstner1998, Niehaus2001} the density functional tight binding (DFTB) approach allows for one to overcome some of these inherent scaling limitations with reasonable accuracy. DFTB has been developed to describe both ground and excited state properties, as are related to Kohn-Sham (KS) DFT and to TDDFT, respectively. In this Feature Article, we highlight how DFTB's excited state formulation can be used to describe plasmons from a quantum level of theory. In many ways, it may seem counterintuitive that such a tight-binding method can be used to describe plasmons, which arise from collective excitations of delocalized electronic states. To reconcile some of these ostensible discrepancies, we briefly discuss the overarching theory behind the self-consistent charge-corrected (SCC) DFTB approach, as well as some underlying approximations and limitations. We then situate this field within the broader context of why a quantum description is necessary to describe plasmons for particles below a critical size and to fully describe plasmon-driven chemistry. We particularly focus on real-time methods used to study electron dynamics under applied electric fields and their implementation within the DFTB framework. While sacrificing some level of quantum mechanical accuracy, real-time time-dependent density functional tight binding (RT-TDDFTB) allows for the investigation of electron dynamics for systems with several hundred (to even thousands of) atoms and at timescales into the picosecond regime. The implementation of this theory is then discussed with respect to understanding the optical properties of individual nanoparticles and their hierarchical structures, with applications thereof to photocatalysis.

\section{DFTB is a Highly Parameterized Semiempirical Method}
Despite originating from KS-DFT, DFTB remains fundamentally a semiempirical method that cannot be expected to match the complete accuracy of higher-level approaches. Its theoretical foundation rests on the assumption that electrons are tightly bound to the nucleus; any interactions (\textit{e.g.}, bonding) between atoms are approximated using empirical parameters that originate from a consideration of the ground state energy. Consequently, DFTB is best suited for highly covalent systems such as hydrocarbons, though many studies have also reported satisfactory performance for more delocalized metallic systems, which we emphasize herein.\cite{Liu2022, DAgostino2018} Energetic contributions are categorized through self-consistent charge corrections, short-ranged repulsive interactions, and pretabulated Slater-Koster integrals which describe orbital overlap.\cite{Slater1954} The majority of the derivations are presented in the \textbf{Supporting Information}, though we briefly present the overarching equations governing this theory in the following section. Throughout these derivations, we use atomic units for simplicity.

\subsection{DFTB is a third-order expansion of KS-DFT}
For an interacting many-body system, the Hamiltonian $H$ is represented in operator form as:
\begin{equation}
    \hat{H} = \hat{T}+\hat{V}_{ext}+\hat{V}_{ee}+\hat{V}_{nn}\, ,
    \label{eq1}
\end{equation}
where $T$ is the kinetic energy, $V_{ext}$ is the energy from an external field (e.g. electron-nuclear contributions), while $V_{ee}$ and $V_{nn}$ constitute interelectronic and internuclear interactions. KS-DFT transforms this complex interacting system into a fictitious non-interacting problem by fixing the electron density $\rho(\mathbf{r})$. This transformation ensures that the noninteracting system's density remains identical to its interacting counterpart. Consequently, for an electron coordinate $\mathbf{r}$, the energy functional for the Hamiltonian in Eq. \ref{eq1} is given by
\begin{align}
  \label{eq2}
    E[\rho(\mathbf{r})] \,= \,&{}T_{s} + \int v_{ext}\rho(\mathbf{r})\,d\mathbf{r} + \frac{1}{2}\iint\frac{\rho(\mathbf{r})\rho(\mathbf{r'})}{|\mathbf{r}-\mathbf{r'}|}\,d\mathbf{r}\,d\mathbf{r'}\\
    &+ E_{xc} + E_{nn}.\nonumber  
\end{align}
In Eq. \ref{eq2}, $T_{s}$ represents the kinetic energy of the noninteracting system, $v_{ext}$ is the one-body external potential, while the third term is the classical Coulomb (or Hartree) energy. Moreover, the exchange-correlation energy term $E_{xc}$ (and its corresponding potential $V_{xc}$) accounts for additional quantum mechanical interactions not explicitly described by classical electrostatic potentials (alongside deviations in kinetic energy between the (non)interacting systems).\cite{Kohn1965} The final term $E_{nn}$ denotes the energy corresponding to internuclear interactions. The energy density functional $E[\rho(\mathbf{r})]$ is then variationally minimized with respect to the orbital wavefunction $\psi_{i}[\rho]$ to yield the fictitious, noninteracting KS wavefunctions. Writing the corresponding single particle equations then gives:

\begin{equation}
\bigg[\frac{1}{2}\nabla^2+v_{eff}(\mathbf{r})\bigg]\psi_i(\mathbf{r}) = \varepsilon_i\psi_i(\mathbf{r}),\quad v_{eff} = v_{ext}+v_{H}+v_{xc}.
\label{eq2.5}
\end{equation}
Consequently, explicitly writing out each term for Eq. (\ref{eq2}) gives
\begin{align}
    E[\rho(\mathbf{r})] = & \sum_{i}n_{i}\varepsilon_i-\frac{1}{2}\iint \frac{\rho(\mathbf{r})\rho(\mathbf{r'})}{|\mathbf{r}-\mathbf{r'}|}\,d\mathbf{r}\,d\mathbf{r'} \nonumber \\ 
    &-\int V_{xc}(\mathbf{r})\rho(\mathbf{r})\,d\mathbf{r} + E_{xc}[\rho]+E_{nn}.
\label{eq3}
\end{align}
The prefactor $n_i$ in Eq. \ref{eq3} denotes the electronic occupation of state $\psi_i$ where $\rho(\mathbf{r}) = \sum_{i}n_{i}|\psi_{i}(\mathbf{r})|^{2}$. This is often taken from the Fermi distribution function $n_i = 2\cdot [\exp(\varepsilon_i-\mu)/k_B T+1]^{-1}$; the chemical potential $\mu$ is chosen such that $\sum_i n_i = N$ with $N = \int \rho(\mathbf{r}) d\mathbf{r}$ yields the total number of electrons. The Hartree and exchange-correlation terms are now subtracted to avoid double counting by virtue of Eq. \ref{eq2.5}.

DFTB is ultimately derived from a Taylor expansion of Eq. \ref{eq4} around a properly chosen KS reference density. In contrast to KS-DFT, which aims to find an electron density which minimizes the total energy, DFTB only assumes a reference density $\rho_0$ where perturbations corresponding to charge density fluctuations (\textit{i.e.}, $\rho(\mathbf{r}) = \rho_0(\mathbf{r}) + \delta\rho(\mathbf{r})$) determine the ground state energy. An appropriately chosen reference density, for example, would be a superposition of constituent neutral atomic electron densities. The total energy from Eq. \ref{eq3} is then expanded up to third order whereby
\begin{align}
    E^{\rm DFTB3}[\rho_0 +\delta\rho] &{}=  E^{(0)}[\rho_0] + E^{(1)}[\rho_0,\delta\rho]  \nonumber \\
    &+ E^{(2)}[\rho_0,(\delta\rho)^2] + E^{(3)}[\rho_0,(\delta\rho)^3] 
    \label{eq4}
\end{align}
where for nuclei with charge $Z_{i}$ and position $\mathbf{R}$,
\begin{equation}
\begin{aligned}
    & \begin{aligned}
    E^{(0)}[\rho_0] = & \frac{1}{2} \sum_{AB} \frac{Z_A Z_B}{|\mathbf{R}_A - \mathbf{R}_B|} 
    - \frac{1}{2} \iint \frac{\rho_0(\mathbf{r}) \rho_0(\mathbf{r'})}{|\mathbf{r} - \mathbf{r'}|} \, d\mathbf{r} \, d\mathbf{r'} \\
    & - \int V_{xc}(\mathbf{r}) \rho_0(\mathbf{r}) \, d\mathbf{r} + E_{xc}[\rho_0(\mathbf{r})],
    \end{aligned} \\
    & \begin{aligned}
    E^{(1)}[\rho_0, \delta \rho] = & \sum_i n_i \langle \psi_i | \hat{H}[\rho_0(\mathbf{r})] | \psi_i \rangle, 
    \end{aligned} \\
    & \begin{aligned}
    E^{(2)}[\rho_0, (\delta \rho)^2] = & \frac{1}{2} \iint \left( \frac{1}{|\mathbf{r} - \mathbf{r'}|} 
    + \frac{\partial^2 E_{xc}[\rho]}{\partial \rho(\mathbf{r}) \partial \rho(\mathbf{r'})} \bigg|_{\rho_0} \right) \\
    & \times \delta \rho(\mathbf{r}) \delta \rho(\mathbf{r'}) \, d\mathbf{r} \, d\mathbf{r'},
    \end{aligned} \\
    & \begin{aligned}
    E^{(3)}[\rho_0, (\delta \rho)^3] = & \frac{1}{3!} \iiint \frac{\partial^3 E_{xc}[\rho]}{\partial \rho(\mathbf{r}) \partial \rho(\mathbf{r'}) \partial \rho(\mathbf{r''})} \bigg|_{\rho_0} \\
    & \times \delta \rho(\mathbf{r}) \delta \rho(\mathbf{r'}) \delta \rho(\mathbf{r''}) \, d\mathbf{r} \, d\mathbf{r'} \, d\mathbf{r''}.
    \end{aligned}
\end{aligned}
\label{eq5}
\end{equation}
The zero- and first-order terms constitute the original DFTB formulation (DFTB1) and is otherwise known as non-SCC DFTB.\cite{Seifert1986} It is important to note that the first order term does not actually depend on $\delta\rho$ as its corresponding terms cancel out during the derivation. DFTB1's non-self-consistent nature requires only a single solution of the Kohn-Sham equations, yielding computational speeds 5-10 times faster than subsequent self-consistent versions. \cite{Elstner2014} Moreover, DFTB1 is best suited for systems with significant covalent character and minimal charge transfer between atoms; such systems include homonuclear systems or molecules with similar atomic electronegativities. Additionally, DFTB1 can describe systems with \emph{complete} charge transfer, such as NaCl.\cite{Elstner2007} However, simply using this first-order approximation has its limitations when modeling systems where many-body interactions significantly influence charge density fluctuations. Later iterations of DFTB (DFTB2 and DFTB3) address these limitations by incorporating higher-order terms to account for atomic charge fluctuations. The second- and third-order terms, represented by two- and three-body integrals, become crucial to accurately describe ionic and metallic (and, by extension, plasmonic) systems.

The total energy expression in Eq. \ref{eq4} can be reorganized into three fundamental contributions:
\begin{equation}
    E^{\rm DFTB3}[\rho_0+\delta\rho] = E_{orb} + E_{\rm SCC} + E_{rep}
    \label{eq6}
\end{equation}
In Eq. \ref{eq6}, $E_{orb} = E^{(1)}$ is the orbital (or band) energy and is calculated as the sum of atomic orbital eigenvalues. The second term, $E_{\rm SCC} = E^{(2)} + E^{(3)}$, describes second- and third-order self-consistent corrections that arise from charge fluctuations. The final term, $E_{rep} = E^{(0)}$ accounts for all other energetic contributions, internuclear repulsion and more cumbersome exchange-correlation effects. The strength of DFTB ultimately lies in its reliance on empirical approximations of the total energy. This highly parametrized approach enables it to achieve results comparable to KS-DFT with wall times less than 0.2\% of the original method for a given system of interest.\cite{Elstner2014}
\subsection{Parameters are Tabulated in Slater-Koster Files}
The DFTB3 Hamiltonian matrix is altogether expressed as a multi-order expansion that progressively captures more detailed electronic interactions, where for atoms $A$ and $B$ containing orbitals $\mu$ and $\nu$, respectively,

\begin{equation}
\begin{aligned}
    \mathbb{H}_{\mu\nu} &= \mathbb{H}_{\mu\nu}^{(0)} + \mathbb{H}^{(2)}_{\mu\nu}[\gamma^{h},\Delta q] + \mathbb{H}^{(3)}_{\mu\nu}[\Gamma,\Delta q],\quad \mu\in A,\nu\in B \\
    \mathbb{H}_{\mu\nu}^{(0)} &= \left\langle \phi_\mu \left|\hat{H}[\rho_0] + \frac{\partial E_{AB}^{rep}}{\partial\rho}\bigg|_{\rho_0} \right|\phi_\nu\right\rangle\\
    \mathbb{H}^{(2)}_{\mu\nu} &= \frac{\mathbb{S_{\mu\nu}}}{2}\sum_{C}(\gamma_{BC}^{h}+\gamma_{AC}^{h})\Delta q_{C}\\
    \mathbb{H}^{(3)}_{\mu\nu} &= \mathbb{S_{\mu\nu}}\sum_{C}\bigg(\frac{1}{3}(\Delta q_{A}\Gamma_{AC}+\Delta q_B \Gamma_{BC}) \\
    &\hphantom{= \mathbb{S_{\mu\nu}}\sum_{C}\bigg(\frac{1}{3}\Delta q_{A}} +\frac{\Delta q_{C}}{6}(\Gamma_{AC} + \Gamma_{BC})\bigg)\Delta q_{C}.
\end{aligned}
\label{eq25}
\end{equation}

 Each order incorporates additional nuances in electronic structure -- $\mathbb{H}_{\mu\nu}^{(0)}$ represents the zeroth-order Hamiltonian based on the reference density (\textit{i.e.}, the orbital and repulsive energies), while the subsequent terms $\mathbb{H}_{\mu\nu}^{(2)}$ and $\mathbb{H}_{\mu\nu}^{(3)}$ progressively incorporate charge fluctuation and chemical hardness effects. The key components of this expansion involve $\mathbb{S}_{\mu\nu}$, the overlap matrix between orbitals $\phi_\mu$ and $\phi_\nu$; $\Delta q$, charge fluctuations; $\gamma^{h}$, a modified interatomic charge density that captures chemical hardness; and $\Gamma$, which captures changes in chemical hardness arising from more nuanced electronic interactions. This Hamiltonian matrix serves as the foundation for solving the corresponding secular DFTB equations from Eq. S4. 
 
 Importantly, both $\mathbb{H}_{\mu\nu}^{(0)}$ and $\mathbb{S}_{\mu\nu}$ are pre-computed once an appropriate pseudoatomic Slater-type orbital basis set is determined, alongside the parameters $\gamma^{h}_{AB}$ and $E^{rep}_{AB}$. For each pair of elements, these parameters are systematically tabulated in Slater-Koster files, eliminating their need to be computed during the DFTB runtime (further discussed in the \textbf{Supporting Information}). This high degree of parametrization underscores DFTB's computational efficiency: DFTB2 is at least 250 times faster than RI-PBE and more than 1000 times faster than B3LYP.\cite{Elstner2014} The use of a minimal basis set alone (\textit{q.v.} \textbf{Supporting Information}) accelerates computation by a factor of 27 compared to KS-DFT, while integral tabulation further improves computational speed by 10 to 40 fold.\cite{Elstner2014} However, matrix diagonalization scales cubically with the number of orbitals ($\mathcal{O}(N^{3})$) and remains the time-limiting step in DFTB.
 \subsection{Real-Time Electron Dynamics Describes Plasmons under Strong Fields}
Excited state phenomena, in contrast to ground state formalisms, require a description of the DFTB Hamiltonian within the time or frequency domain. In this regard, time-dependent density functional theory (TDDFT) \cite{runge84} is the most popular method to compute the excitation energies of many-electron systems due to its reasonable tradeoff between accuracy and efficiency \cite{casida12, adamo13}. The most widely implemented version is derived from linear response theory (LR-TDDFT) per the Casida formalism:\cite{casida95} 
\begin{equation}
    \Omega\mathbf{F}_i = \omega_i^{2}\mathbf{F}_i,
\end{equation}
where $\Omega$ is the response matrix whose elements depend on the ground state KS states and the exchange-correlation kernel chosen, $\omega_i$ are the frequencies corresponding to the vertical excitation energies, and the eigenvectors $\mathbf{F}_i$ contain all the information needed to derive excited state properties (\textit{e.g.} total spin and transition oscillator strengths). The former two terms result from the density-density response matrix $\chi(\mathbf{r},\mathbf{r'},\omega)$ which describes how the electron density at $\mathbf{r}$ responds to a perturbation at $\mathbf{r'}$ as a function of frequency $\omega$. The full spectrum, which includes the orbital energy differences $\omega_i$, affects how transitions are weighted and how the response matrix elements $\Omega$ are constructed.\cite{petersilka96}. Note that in DFTB, only valence excited states can be computed due to the minimal basis set employed. The polarizability tensor $\overset{\leftrightarrow}{\alpha}(\omega)$ can be subsequently computed (within the Kramers-Heisenberg dispersion relation) by
\begin{equation}
    \overset{\leftrightarrow}{\alpha}(\omega) = \sum_{i \neq 0}\frac{e^{2}f_{i}}{m_{e}(\omega_{i}^{2}-\omega^{2})},
\end{equation}
where the oscillator strengths $f_{i}$ are obtained from the eigenvectors $\mathbf{F}_i$ and $m_e$ and $e$ is the electron rest mass and charge, respectively.\cite{casida95} 

While LR-TDDFT is capable of studying a few low-lying excited states of an isolated molecule, it becomes prohibitively expensive in both memory and computational time when analyzing broadband spectra \cite{bruner16} or for systems with a high density of states. This is often the case for semiconductors\cite{Ezra2024} and many plasmonic systems (\textit{e.g.} Al, alkali metals), which are characterized by free-electron behavior with few interband transitions. \cite{Bae2022, Sundararaman_2014} Importantly, the linear Casida equation is iteratively solved within a basis spanning (occupied) $\times$ (virtual) orbitals, making LR-TDDFT calculations for these systems inaccessible even at the DFTB level of theory.\cite{Sinha-Roy2018,Rossi2017} Such studies are often performed using smaller noble-metal plasmonic nanostructures and extrapolated to experimentally-observed sizes.\cite{DAgostino2018} Another strategy to partially mitigate this issue involves approximating the two-center electron integrals involved in calculating both $\Omega$ and $\mathbf{F}_i$.\cite{wang24} Nonetheless, TDDFT calculations employing (semi)local functionals lead to spurious low-lying charge transfer states, a well-known issue underlying these exchange-correlation functionals.\cite{Koppen_2012, Silverstein_2010} As DFTB is parameterized using semilocal exchange-correlation functionals (in order to properly map Eq. \ref{eq3} to \ref{eq6}), these errors are inherited in TDDFTB calculations as well. More recently, long-range corrected DFTB has been developed to reconcile some of these discrepancies and exhibits similar accuracy to range-separated DFT methods at a significantly reduced computational cost,\cite{Vuong_2018, vanderHeide_2023} though specific parameterizations for plasmonic metals remain unavailable as of yet.

Alternatively, real-time TDDFT (RT-TDDFT) can be used to directly solve the time-dependent Kohn–Sham equation \cite{Yabana1996} (or its density matrix-based analogue, the Liouville-von Neumann equation of motion).\cite{watanabe02,li05} This non-perturbative approach is capable of capturing nonlinear photophysical effects and allows for one to generate broadband spectra. The spectral range can be extended to lower wavelengths by propagating the system over longer times, and its resolution can be improved by decreasing the timestep. This presents a significant departure from iterative LR-TDDFT methods that explore excited states by gradually expanding the excitation space.

Solving the time-dependent KS equation requires the use of a full time-dependent propagator, which can be expressed in terms of the time-ordered evolution operator:
\begin{equation}
\hat{U}(t,0) = \mathcal{T}\exp\bigg\{-{\rm i}\int_0^{t}\hat{H}(\tau)\,d\tau\bigg\},
\label{eq31}
\end{equation}
where $\mathcal{T}\exp$ is a shorthand notation for the time-ordered exponential.\cite{castro04} However, due to the complexity of this operator, it must be approximated. Several numerical methods have been developed to do so, including the exponential midpoint (and modified midpoint) algorithm, the enforced time-reversal symmetry method, Magnus expansion, and split-operator method.\cite{castro04,gomez18} Consequently, the accuracy and efficiency of RT-TDDFT calculations heavily depend on the approximations used for the time-dependent propagator and the spectral properties of the time-dependent Hamiltonian.\cite{ghosal22}

Within the tight-binding (RT-TDDFTB) formalism, a classical time-dependent propagator is used to approximate $\hat{H}(\tau)$ at the DFTB level, allowing for one to study larger plasmonic systems over extended timescales ($\sim$ hundreds of fs to ps). Nuclear dynamics are treated semiclassically,\cite{Hourahine2020} while electronic states are allowed to evolve under an applied field; coupling between the two is captured \textit{via} the Ehrenfest model.\cite{Bonafe2020} In this approximation, the nuclei are treated as classical particles subject to a force derived from a weighted average of all the electronic states of the system.\cite{Ehrenfest1927} Specifically, the density matrix $\rho_{\mu\nu} = \sum_{i}n_{i}(c_{\mu i} c_{\nu i}^{*})$ is propagated according to the Liouville-von Neumann equation of motion
\begin{equation}
\frac{\partial \rho}{\partial t} = -{\rm i} \left( \mathbb{S}^{-1} \mathbb{H} \rho - \rho \mathbb{H} \mathbb{S}^{-1} \right)- \left( \mathbb{S}^{-1} \mathbb{D} \rho + \rho \mathbb{D}^{\dagger} \mathbb{S}^{-1} \right)
\label{eq32}
\end{equation}
where $\mathbb{H}$ is now time dependent by virtue of the applied electric field. 
The density matrix is evaluated by integrating Eq. 12 for a given initial ($t\=0$) density matrix $\rho_0$ representing the ground state of the system. The second term in the above equation accounts for dissipation and energy exchange between electrons and nuclei. The nonadiabatic coupling matrix, $\mathbb{D}_{\mu\nu} = \frac{\partial \mathbf{R}_{B}}{\partial t}\cdot\nabla_B\mathbb{S}_{\mu\nu} = \langle \phi_{\mu}|\frac{\partial\phi_{\nu}}{\partial t}\rangle$ introduces a mechanism by which nuclear motion can induce electronic transitions, leading to phenomena such as electronic thermalization (key in plasmon-generated hot carrier dynamics,\cite{Lin2009, Brongersma2015, Hartland2017, Wu2022} though with significant limitations; \textit{vide infra}). The equation of motion is typically integrated using a leapfrog scheme where the density matrix at time $t_{i+1}$ is obtained from its value at time $t_{i-1}$ and its derivative at time $t_{i}$:
 \begin{equation}
     \rho(t_{i+1}) = \rho(t_{i-1})+2\Delta t\frac{\partial\rho}{\partial t}\bigg|_{t_{i}}
 \end{equation}
A major challenge for real-time methods is the extremely short timestep (often on the attosecond scale) necessary to accurately capture electron dynamics. In contrast, ground-state \textit{ab initio} molecular dynamics simulations (which do not involve electron propagation) typically use timesteps between 0.5 and 1 fs. Larger timesteps may be possible \textit{via} an optimal gauge choice, enabling propagation on the order of 10-100 attoseconds,\cite{Jia2018} although this remains an active area of research.

The forces on each atom are computed using the Hellmann-Feynman theorem, where the time-evolved density matrix is averaged over the distribution of electronic states such that
 \begin{equation}
     m_{A}\frac{\partial^{2}\mathbf{R}_{A}}{\partial t^{2}} = -\nabla_{A}\bigg[\frac{\text{Tr}(\rho\mathbb{H})}{\text{Tr}(\rho)}\bigg]\,,
 \end{equation}
where $m_{A}$ denotes the mass of atom $A$. The atomic charges are then calculated by the Mulliken approximation, $q_{A} = \text{Tr}_{A}[\rho\mathbb{S}]$.\cite{Bonafe2020, Hourahine2020} The nuclear and electronic systems interact simply through expectation values and thus this approach is a mean-field approximation to the actual coupled electron-nuclear dynamics. This approximation may break down under circumstances where electron-nuclear correlations become important, such as at the crossing of conical intersections.
 
In the weak-field limit, the results of RT-TDDFT and LR-TDDFT should agree. Nevertheless, RT-TDDFTB encounters notable challenges when describing phenomena driven by intense external fields, such as above-threshold ionization\cite{schafer93} multi-photon processes,\cite{sakai03} charge-resonance enhanced ionization,\cite{bocharova11} and high harmonic generation.\cite{chu01} Specifically, two major sources of error, aside from limitations inherent to using a minimal basis set, include an inadequate description of excited state potential energy surfaces and the long-range behavior of the Coulomb potential. These issues complicate the accessibility of high-energy KS states during time propagation. Even in the weak-field limit, Ehrenfest dynamics does not obey the principle of detailed balance and can perform poorly when a system has multiple equiprobable states where the potential surfaces are significantly different.\cite{Parandekar2006,Provorse2016} However, Ehrenfest dynamics is expected to perform well over short time scales in systems with many similar potential energy surfaces (as is often the case for plasmonic metallic nanoparticles) since it restricts motion to a single average potential energy surface.

RT-TDDFTB can be used to calculate electronic absorption spectra. Following a weak delta function electric field impulse at $t = 0$, the system evolves without any external perturbation. The spectrum is calculated from the frequency-dependent induced dipole moment $\vec{\mu}(\omega)$, related to the field $\mathbf{E}(\omega)$ \textit{via} the polarizability tensor \cite{Yabana1996}
 \begin{equation}
     \vec{\mu}(\omega) = \overset{\leftrightarrow}{\alpha}(\omega)\mathbf{E}(\omega).
     \label{eq_pol}
 \end{equation}
In practice, a damping factor is often applied to smooth the Fourier transformed spectra, thereby reducing ringing (\textit{i.e.}, rapid oscillations) and providing a more realistic description of excited state dynamics (\textit{e.g.} finite lifetime effects on the plasmon linewidth and chemical interface damping\cite{Douglas-Gallardo2016}). The absorption spectrum $\sigma(\omega)$ is then calculated from the frequency-dependent polarizability tensor as
 \begin{equation}
     \sigma(\omega) = \frac{4\pi\omega}{3c}\im[\text{Tr}(\overset{\leftrightarrow}{\alpha}(\omega))].
     \label{eq36}
 \end{equation}
The maximum peak in the absorption spectrum typically corresponds to the plasmon mode, and this resonant frequency can be used in subsequent calculations, such as laser-driven dissociation.\cite{Herring2023,Giri2023,Li2023,Zhang2018} In the following sections, we discuss some practical implementations of this theory, focusing on its performance and limitations when describing the unusual optical properties of small metal clusters, alongside nonlinear spectroscopy and photocatalysis.
 
\section{DFTB Enables a Reconciliation between Classical and Quantum Plasmonics}
Large metal nanoparticles ($\sim$10-200 nm) behave like conventional electrical conductors and their plasmon resonances can be understood within a classical picture. Atomically-precise nanoclusters, however, reveal an entirely different world. Nonlocal effects (where the frequency-dependent dielectric function additionally depends on the incident wavevector) begin to predominate as the electron mean free path becomes commensurate with the particle size ($\approx$ 5 nm for Al and 50 nm for Ag), leading to anomalous plasmon broadening.\cite{Schatz1983, Kanter1970, McMahon2009} As the particle size approaches the Fermi wavelength ($\approx$ 0.2 to 0.5 nm for Al and Ag), confinement of electron motion transforms its otherwise continuous density of states into discrete energy levels.\cite{Blackman2018,Galchenko2019} These tiny clusters behave more like molecules than metals, with sharp electronic transitions replacing their broad plasmon bands.\cite{Zhu2008} It then follows that their unusual optoelectronic properties require a quantum description rather than the classical picture typically used. Kubo showed that the spacing between energy levels scales inversely with particle volume.\cite{Kubo1962,Kawabata1966} This can be understood through a free electron model where the energy gap – or eponymous Kubo gap $\delta$ – near the Fermi energy $E_{F}$ is given by
\begin{equation}
    \delta = \frac{4E_F}{3N},
\end{equation}
where $N$ is the number of valence electrons. The particle's electronic behavior is determined by the relationship between the Kubo gap and thermal energy fluctuations. When $\delta$ falls below the thermal energy $k_B T$, the particle will be metallic; if above, insulating. For example, at room temperature the heuristic number of valence electrons for $\delta$ to equal $k_B T$ is around 287 for Ag ($E_{F}$ = 5.51 eV) and 60 for Al ($E_{F}$ = 11.6 eV). For Ag this roughly corresponds to a 2 nm nanoparticle, though for Al this is strikingly smaller – only about 0.4 nm (approximately 20 atoms). Classical approaches tend to address these scale-dependent variations by empirically modifying the dielectric function to include enhanced electron-surface scattering in small particles,\cite{Kreibig2013, Fan2014, Ross2015} or by using a nonlocal dielectric function outright.\cite{McMahon2009-2, Mortensen2014} Though at this scale, any nuances in electronic structure are dominated by quantum size effects that dictate any emergent physicochemical properties.\cite{Mandal2013,El-Sayed2001} 

To this end, DFTB allows for one to calculate the optical properties of ultrasmall metal nanoparticles from a purely quantum mechanical standpoint. While DFTB parameterizations rely on generalized gradient approximation (GGA) functionals primarily designed primarily for ground state properties, these semilocal functionals can still reasonably describe excited state phemonena.\cite{aikens2008, Varas2016} The approximations used at this level of theory ostensibly capture the essential screening effects that influence plasmon behavior (particularly in \textit{d}-electron systems;\cite{He2010} \textit{vide infra}), even if they do not fully account for optical nonlocalities. Importantly, nonlocal dielectric functions, which account for spatial variations in a material's response to an external field, less markedly affect absorption (which is directly calculable from RT-TDDFTB per Eq. \ref{eq36}). This is largely because absorption (which largely depends on the optically-allowed density of states) predominates at sub-10 nm length scales, rather than scattering processes which depend on the field's detailed spatial distribution.

Salient limitations nonetheless remain where an inadequate parameterization of excited state interactions lead to spurious low-lying charge transfer states -- a well-known limitation of GGA functionals at the KS-DFT level, delocalization error notwithstanding.\cite{Bryenton_2023} Moreover, the intense absorption band generally ascribed to plasmons has been shown to have strong interband character when calculated with GGA functionals.\cite{Seveur2023} Although range-separated DFTB\cite{Vuong_2018} should better represent the long-ranged Coulomb potential necessary to describe plasmons,\cite{Silverstein_2010, Franck_2013, Seveur2023} optical spectra are predicted to manifest as sparse transitions for smaller clusters (\textit{cf.} that computed at the KS-DFT level in Ref. \citenum{Wu_2023}) rather than the dense manifold of higher-energy excited states necessary for plasmon-like transitions in metallic nanoparticles. Consequently, plasmon-like transitions would only appear for much larger particles (which DFTB can nonetheless simulate). The use of range-separated dielectric-dependent corrections, as has been recently implemented for periodic systems at the KS-DFT level\cite{Refaley_2015, Skone_2016} has been shown to more realistically account for Coulombic screening. Extensions thereof to DFTB could offer a promising means to describe the delocalized transitions and charge transfer excitations characteristic of plasmons.

\begin{figure}[h!]
    \centering
    \includegraphics[width=0.7\linewidth]{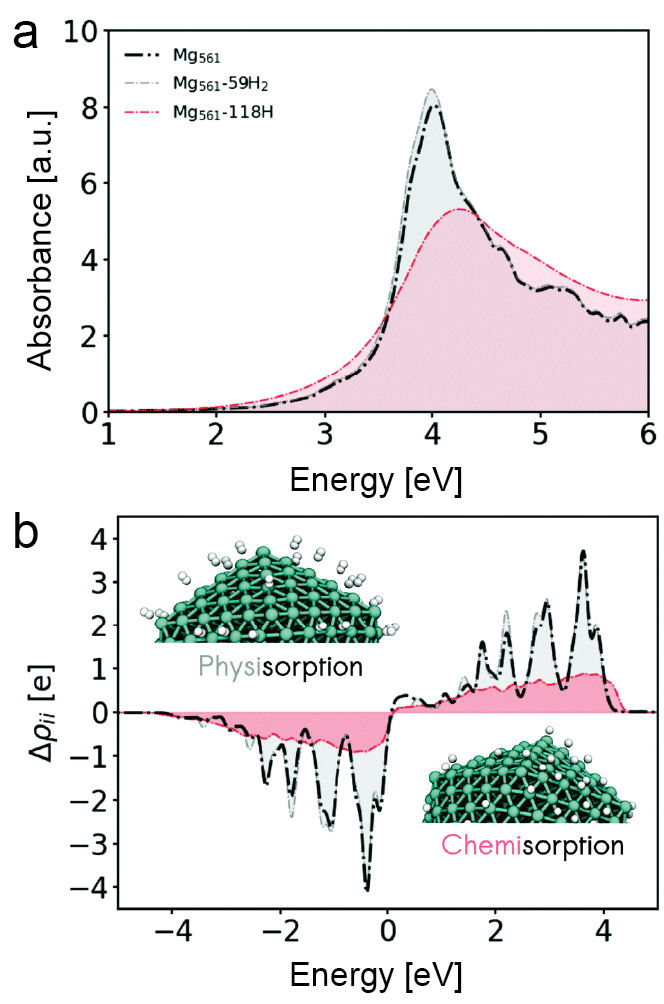}
    \caption{Hydrogen absorption dampens the plasmonic response of Mg nanoclusters. \textbf{(a)} RT-TDDFTB-calculated optical absorption spectra of \ch{Mg561}, with 59 \ch{H2} molecules physisorbed onto and 118 H atoms absorbed \textit{into} the \ch{Mg561} cluster. \textbf{(b)} Corresponding electronic energy distribution, calculated as the difference in electron populations projected onto the ground state orbitals. Reproduced with permission by Ref.\citenum{DouglasGallardo2021}. \textcopyright\, 2021 Royal Society of Chemistry.}
    \label{Mg_plasmonic}
\end{figure}

While DFTB has been parametrized for some main group plasmonic materials like Al, Na, and K,\cite{Kubillus2015} few studies exist that systematically examine their plasmonic properties. Douglas-Gallardo \textit{et al.} investigated the plasmonic properties of Al nanoparticles \emph{via} a combination of molecular dynamics and RT-TDDFTB.\cite{Douglas-Gallardo2017} The authors report that incorporating $d$ orbital angular momentum terms in the Hamiltonian matrix (from Eq. S8) is necessary to describe their optical and structural properties. Oxidation was found to gradually separate their otherwise continuous density of states, resulting in attenuated electron dynamics and shortened plasmon lifetimes. Similarly, we have recently calculated the optical absorption spectra for Al nanoparticles where, due to Al's free electron character, particles were found to display strong symmetry-dependent plasmon modes.\cite{Chellam_2025} A related study found through DFTB that Mg nanoclusters can produce hot carriers with energies up to 4 eV (\textbf{Fig. \ref{Mg_plasmonic}}).\cite{DouglasGallardo2021} These generated hot carriers energetically align with the electronic states of physisorbed \ch{H2} molecules, which can then promote their dissociation. Chemisorbed (\textit{i.e.}, absorbed) \ch{H2}, conversely, was found to lead to strong chemical interface damping, thereby reducing the plasmon lifetime and blueshifting the plasmon wavelength. This interaction resulted in a diminished hot electron population, though the remaining electrons possessed slightly higher energies. The paucity of such studies suggests a rich avenue to be explored, particularly given these metals' Earth abundance and free electron character.

Despite these promising applications to main group metals, DFTB's inherent tight-binding construction limits its ability to accurately model the anomalous plasmonic properties exhibited by small alkali metal particles. At these sizes, both quantum effects and detailed surface interactions become important as electrons interact strongly with the surface; one aspect includes the spillover of conduction electrons at the particle surface, which would effect a red shift with decreasing particle size.\cite{Selby1989,Selby1991,deHeer1993,deHeer1987} As an analytical reference, the optical response of a spherical metal nanoparticle much smaller than the wavelength of irradiated light (\textit{i.e.}, below the quasistatic limit) can be described by the dipole polarizability $\alpha(\omega)$ that features a nonlocal, hydrodynamic extension of the Clausius-Mossotti relation,\cite{Schatz2003} given by\cite{Raza2013}
\begin{align}
    \alpha(\omega) & = 4\pi R^{3}\frac{\varepsilon_D - \varepsilon_b(1+\delta_{nl})}{\varepsilon_D + 2\varepsilon_b(1+\delta_{nl})}, \nonumber \\
    \delta_{nl} &= (\varepsilon_D - 1)\frac{j_1(k_{nl}R)}{k_{nl}R\,j_1'(k_{nl}R)},
    \label{eq38}
\end{align}
where $\varepsilon_D = 1-\frac{\omega_p^{2}}{\omega^{2}+i\gamma\omega}$ is the Drude dielectric function, $k_{nl}^2 = \frac{\omega^2+i\omega\gamma-\omega_p^2}{\beta^2+\mathcal{D}\gamma-i\mathcal{D}\omega}$ is the longitudinal wavevector, and $j_1$ is the first order spherical Bessel function of the first kind. From Thomas-Fermi theory, $\beta = \sqrt{\frac{3}{5} v_F}$ is a characteristic parameter associated with pressure waves in the electron gas (where $v_F$ is the Fermi velocity), and $\mathcal{D}$ is the characteristic charge-carrier diffusion coefficient. Evaluating for the pole of Eq. \ref{eq38} yields the complex-valued resonance frequency $\omega = \omega'+i\omega''$, which is semiclassically related to the bulk plasma frequency $\omega_p$ and cluster radius $R$ as\cite{Mortensen2014}
\begin{align}
    \omega' &{} = \frac{\omega_p}{\sqrt{3}}+\frac{\beta\sqrt{2}}{2R}\\
    \omega'' & =  -\frac{\gamma}{2}-\frac{\sqrt{6}}{24}\frac{\mathcal{D}\omega_p}{\beta R},
    \label{eq39}
\end{align}
where the real part $\omega'$ gives the plasmon resonant frequency, while the imaginary component $\omega''$ relates to the plasmon damping rate $\gamma$ and hydrodynamic effects (\textit{i.e.}, convective and diffusive charge transport) which spread the charge density and broaden the plasmon linewidth. As the high kinetic energy of the conduction band $s$ electrons in alkali metals cannot be efficiently screened by other electrons comprising the particle's electronic states,\cite{Fan2014} the electron density will consequently extend beyond the nominal surface by some distance $\delta$.\cite{Snider1983,Morton2011,deHeer1993,Tiggesbaumker1993} Accounting for this spill-out effect yields a modified plasma frequency $\omega_{p0}$ dependent on cluster size, such that\cite{Huang2020}
\begin{equation}
    \omega_{p0} \simeq \frac{\omega_p}{(1+\frac{\delta}{r_0})^{\frac{3}{2}}},
    \label{eq40}
\end{equation}
where $r_0 = N^{1/3}r_s$ is the nominal particle radius related to the number of atoms $N$ and the Wigner-Seitz radius $r_s$. Accordingly, this yields $R = r_0 + \delta$. With increasing particle size, the polarizability reaches the classical limit of a perfectly conducting sphere: that is, $\alpha =  \frac{4}{3}\pi r_0^{3}$. Indeed, the polarizability of a sphere in the bulk limit is smaller than that for smaller particles, where valence electron spill-out increases the clusters' effective radius. As the plasmon resonant frequency is inversely proportional to the polarizability, $s$-electron spillover would thus induce a red shift compared to classical Mie theory.\cite{Morton2011}

\begin{figure}[t]
    \centering
    \includegraphics[width=\linewidth]{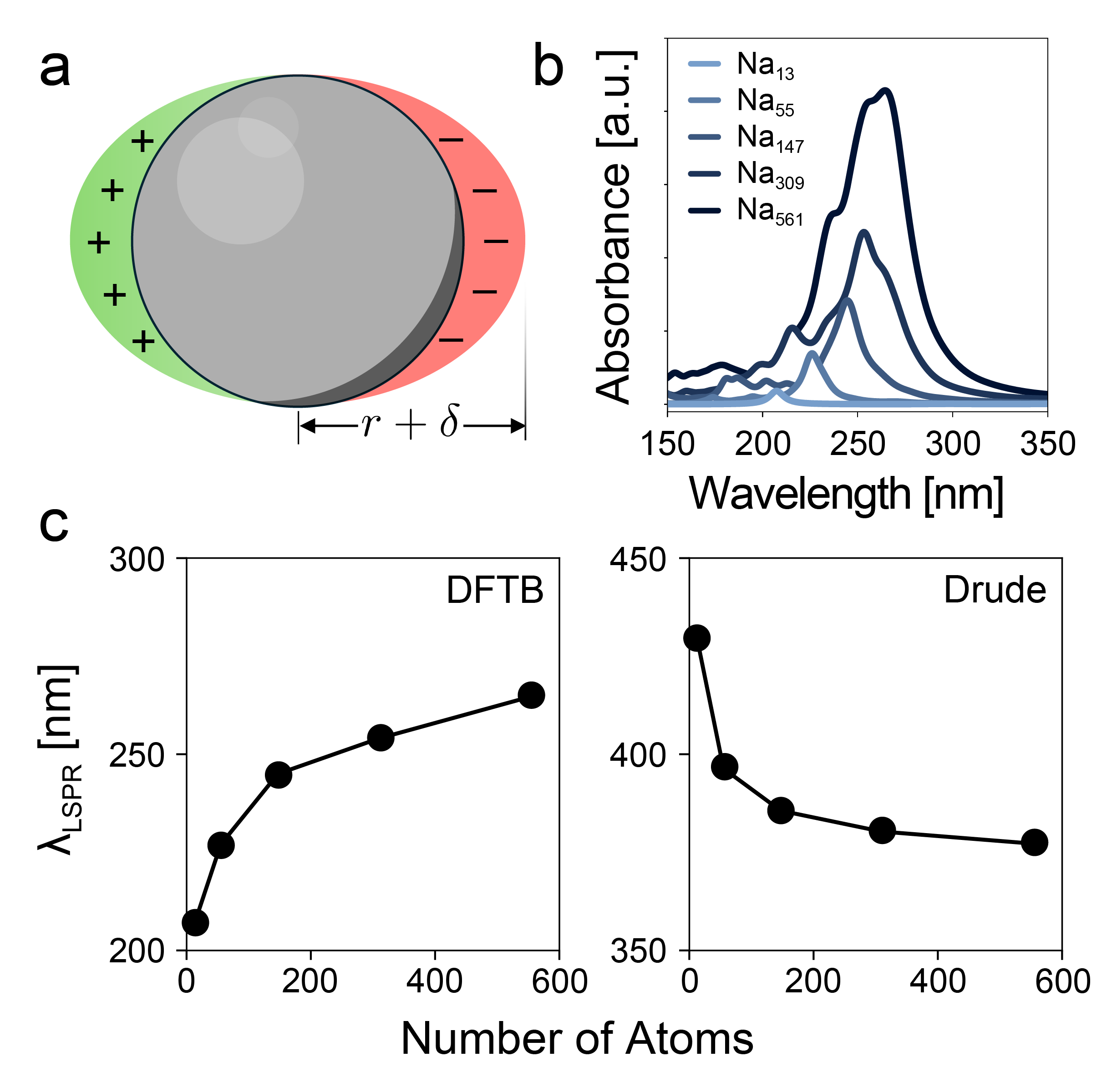}
    \caption{DFTB fails to account for $s$-electron spillover. \textbf{(a)} Schematic of plasmon spillover in alkali metal nanoparticles, where the electron density overextends from a particle of radius $r$ by a distance $\delta$. \textbf{(b)} Calculated absorption spectra for Na clusters spanning \ch{Na13} to \ch{Na561}. Calculations were performed using the {\asciifamily matsci-0-3} parameter set, with a density matrix propagation of 100 fs. \textbf{(c)} Comparison of plasmon peak position with DFTB and the nonlocal hydrodynamic Drude model described by Eqs. \ref{eq39} and \ref{eq40}. The bulk plasma frequency used for Na is 5.89 eV with a damping parameter $\gamma = 0.0276$ eV and Fermi velocity $v_F = 1.07\times 10^6$ m/s, obtained from Ref. \citenum{Blaber_2009}.}
    \label{fig:drude-Na}
\end{figure}

For sodium ($r_{s} = 4$ a.u.), the effective radial increase due to electron spillover is known to be $\delta = 1.3$ a.u. beyond the physical surface.\cite{deHeer1993, Teperik_2013}. We have shown the DFTB-calculated absorption spectra for sodium nanoparticles ranging from \ch{Na13} to \ch{Na561} (\textbf{Fig. \ref{fig:drude-Na}b}) and a comparison to the plasmon peak position \textit{via} the nonlocal hydrodynamic model given by Eqs. \ref{eq39} and \ref{eq40}. Many discrepancies between the two are observable. While both theories show convergence of the plasmon wavelength $\lambda_{LSPR}$ as the particle size increases, they predict contradictory trends (\textbf{Fig. \ref{fig:drude-Na}c}). DFTB shows that these particles absorb in the deep UV (between 200-300 nm) and a concomitant redshift with particle size. On the other hand, the nonlocal hydrodynamic model predicts absorption in the visible and a corresponding blueshift, consistent with reduced $s$ electron spillover in larger particles. Numerous higher-level KS-DFT simulations have examined their photoabsorption spectra,\cite{vanGisbergen2001, Hong2011, Lindgren2015} which run in agreement with the nonlocal hydrodynamic model. However, extending these insights to DFTB is relatively limited, given that parameterization thereof often relies on oxidized states (rather than in zerovalent states).\cite{Frenzel2009,Gaus2013} Additionally, DFTB is inherently limited by artificial orbital compression and the use of a minimal basis set (per Eq. S2); that is, the ``tightly-bound" electrons intrinsic to this theory cannot thus far rigorously describe the otherwise diffuse $s$ states responsible for the spillover effect.

Most literature-reported studies have used DFTB to study noble metal plasmonics, likely due to their popularity in experimentation (a consequence of their chemical stability). While ultrasmall alkali metal and certain \textit{p}-block nanoparticles almost behave like free electron metals, the electronic structure of their noble metal counterparts markedly differs owing to the presence of $d$ electrons: low-lying $d$ electrons screen valence $s$ electron intraband transitions and directly participate in $d\,\rightarrow sp$ interband transitions which dampen the plasmon resonance. Conversely, as the Ag particle size decreases, its surface-to-volume ratio increases, reducing $d$ electron screening, thereby increasing the plasmon frequency. This effect compensates for any spill-out effects conferred by the valence $s$ electrons.\cite{Liebsch1993,Scholl2012,Raza2013} Additionally, the 4$d$ electrons in ultrasmall Ag nanoparticles comprise a polarizable background that screens interactions between valence 5$s$ electrons.\cite{Liebsch1993-2,Jensen2009, Coronado2011, Scholl2012} Consequently, in contrast to the redshift found in alkali metal clusters, these surface-assisted processes induce an anomalous blueshift with decreasing particle size (though it should be noted that the presence of a dielectric ligand shell can effect an opposite trend; \textit{cf.} Ref. \citenum{Peng2010}).\cite{Raza2013,Scholl2012}

\begin{figure}[h]
    \centering
    \includegraphics[width=\linewidth]{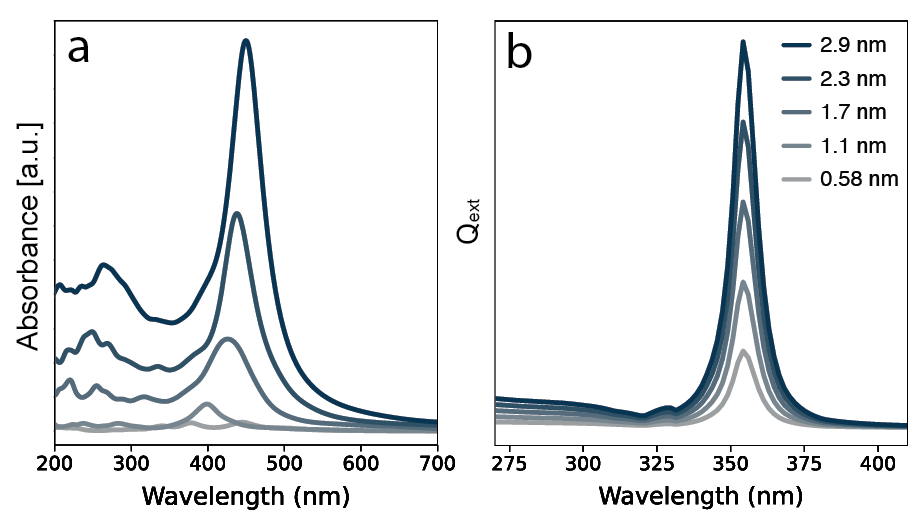}
    \caption{DFTB captures $d$-orbital screening of plasmonic metals with reasonable accuracy. \textbf{(a)} DFTB-calculated linear absorption spectra of icosahedral Ag nanoclusters between 0.58 and 2.9 nm. Calculations were performed using the {\asciifamily hyb-0-2} parameter set, with the density matrix propagated for 50 fs. \textbf{(b)} Mie theory-calculated extinction efficiency for spherical Ag nanoparticles of identical sizes. Calculations were performed using the dielectric function from Johnson and Christy (1972).\cite{Johnson_Christy_1972}}
    \label{fig:Ag-Mie}
\end{figure}

Classical models only account for valence 5$s$ electrons and cannot describe these effects.\cite{Genzel1975, Lindfors2004, Peng2010, Kreibig2013} \textit{Ab initio} methods, on the other hand, explicitly account for $d$ electronic states. Further, as these states are more strongly localized and do not exhibit spill-out effects like $s$ electrons, their contributions are more ostensibly captured by DFTB.\cite{Douglas-Gallardo2016,Giri2023} These results are noticeable in \textbf{Fig. \ref{fig:Ag-Mie}a}, where a blueshift of approximately 50 nm is observed when icosahedral Ag nanoclusters decrease in size from 2.9 to 0.58 nm. This blueshift runs concomitant with an anomalous linewidth broadening due to electron-surface interactions alongside any emergent quantum size effects. In contrast, Mie theory, which does not account for surface scattering and the quantum nature of the dielectric function altogether, displays minimal changes in the plasmon wavelength with spherical particle size \textbf{(Fig. \ref{fig:Ag-Mie}b)}.

Additionally, $d$ electrons more strongly contribute to plasmonic transitions in Au compared to Ag nanoparticles (a consequence of the smaller energy difference between 5$d$ and 6$s$ states).\cite{Aikens2006, Weissker2015} This, in turn, arises from relativistic effects which are essential to include when accurately describing the optical properties of Au clusters. At the DFTB level, the atomic orbital energies and two-center Slater-Koster integrals are obtained from scalar relativistic KS-DFT calculations.\cite{Heera_1984, Koskinen_2006} These effects are effectively accounted for within the Hamiltonian matrix elements (from Eq. S3) without resorting to specific scalar relativistic operators. Thus, while these relativistic effects are implicitly accounted for in $E_{rep}$, neither these effects nor core electronic states (for which relativistic effects predominate) are explicitly described within the DFTB framework, leading to small deviations between multiple levels of theory (and experiment).\cite{Fihey2015,Giri2023,Bae2015}

Quantum models provide a superior description of plasmons at sufficiently small sizes. These methods naturally interpret emergent optical phenomena in terms of discrete, molecular-like transitions. However, as the particle increases in size, the ``collectivity" criterion used to describe plasmons (\textit{i.e.}, as a linear combination of many single-particle excitations) begins to predominate.\cite{Gieseking2016, Gieseking2020, Seveur2023} From an experimental purview, this excitonic to plasmonic transition was identified to be around 2.3 to 1.7 nm for Au clusters.\cite{Zhou2016} At the same time, scalability limitations in KS-DFT render a comprehensive analysis of such excitations and their contributions to the plasmon-like absorption band intractable.\cite{Rossi2017} Prior studies have used symmetry-based arguments to study larger, tetrahedral nanoparticles at the TDDFT level;\cite{Seuret-Hernandez2023, Seveur2023} one study from our group evaluated the optical properties of these structures up to 120 atoms, making it possible to connect TDDFT results with those from classical electrodynamics (\textit{via} the discrete dipole approximation).\cite{aikens2008} Further, when a ground state KS-DFT calculation was performed followed by tight-binding approximations to evaluate their excited state properties, comparable plasmon peak positions were obtained (with \textit{ca.} 0.15 eV deviations) for similarly-sized Au, Ag, and bimetallic alloy tetrahedral particles with only a tenth of the total wall time.\cite{Asadi-Aghbolaghi2020} Even faster runtimes were found to be possible (0.9\% compared to DFT) at the full DFTB level of theory.\cite{DAgostino2018} For significantly larger sizes (up to $10^5$ atoms\cite{Prodan2002,Prodan2003}), the optical properties of nanoparticles can be efficiently calculated using a jellium model, wherein the quantum nature of electrons can be accounted for without an explicit introduction of the constituent lattice. This lends itself possible for free-electron metals with spherical geometries;\cite{Montag1995} though, this theory cannot examine plasmonic systems with strong interband character as well as anisotropic nanoparticles where the spatial charge distribution drastically affects its optical response.\cite{Yin2008,Zhang2018}

\begin{figure}
    \centering
    \includegraphics[width=\linewidth]{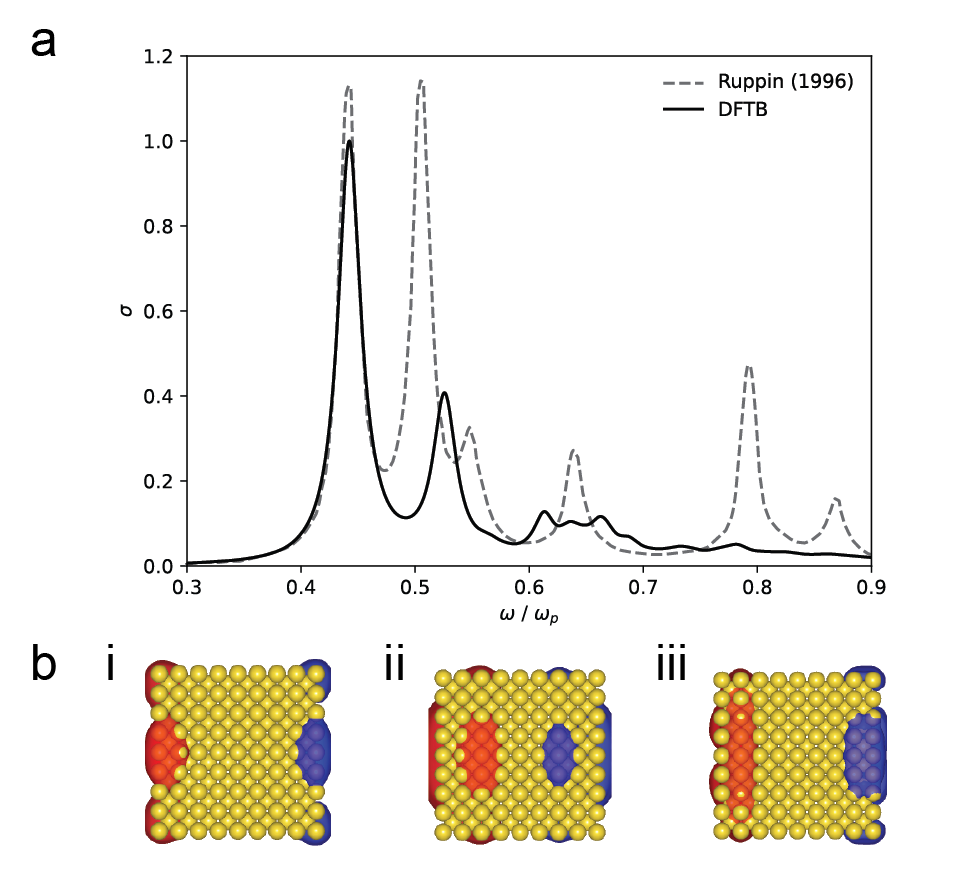}
    \caption{DFTB allows for a comparison to particles in the classical electrodynamics regime. \textbf{(a)} Comparison of DFTB-calculated absorption spectra (using the {\asciifamily matsci-0-3} Slater-Koster parameter set) for a 3 nm \ch{Na1241} cubic particle with that computed by Ref. \citenum{Ruppin_1996}. The frequency of the DFTB-calculated spectrum was scaled to make the first peak coincide with that of the other spectrum. The density matrix was propagated for 100 fs and broadened by 10 fs when dampened. \textbf{(b)} Corresponding DFTB-calculated charge density distributions at $\omega/\omega_p$ = \textbf{(i)} 0.439, \textbf{(ii)} 0.521, and \textbf{(iii)} 0.607. The red and blue shaded regions correspond to positive and negative charge density, respectively.}
    \label{fig2}
\end{figure}

DFTB, at its core, is still an orbital-based method. The use of this theory enables a comprehensive evaluation of this excitonic-to-plasmonic transition with much greater accessible sizes – reported up to 1415 atoms.\cite{Douglas-Gallardo2019} The tractability of DFTB to much larger length scales moreover allows for a direct reconciliation between the quantum and classical regimes. As a point of comparison, Fuchs has shown that the optical properties of metal cubes below the quasistatic limit can be classically expressed as a sum of symmetry-dependent normal modes $m$, The dielectric susceptibility $\chi$ is then given by\cite{Fuchs_1975}
\begin{equation}
    \chi(\omega) = \frac{1}{4\pi}\sum_{m}\frac{C(m)}{(\varepsilon/\varepsilon_b-1)^{-1}+n_m}\, ,
    \label{eq:ruppin}
\end{equation}
where $\varepsilon$ and $\varepsilon_b$ denote the dielectric function of the particle and its surroundings, respectively, and $C(m)$ is the strength of the $m^{th}$ peak which satisfies the sum rule $\sum_mC(m) = 1$. Calculation of the absorption cross section $\sigma = 4\pi(\omega/c)\sqrt{\varepsilon_b}a^{3}\text{Im}[\chi(\omega)]$ for a particle of edge length $a = 3$ nm shows six main resonances with two predominant modes at low energies. The spectrum calculated \textit{via} DFTB shows qualitative agreement with this classical theory. Indeed, we see multiple resonant peaks corresponding to different plasmon symmetries at similar energies to those predicted by Eq. \ref{eq:ruppin} (\textbf{Fig. \ref{fig2}}). We note that neither model accounts for electron spillover, however (\textit{vide supra}). The use of DFTB consequently presents a promising means by which the optical properties of metal nanoparticles can be rigorously examined for much larger particles, allowing for direct comparisons between classical electrodynamics and quantum mechanics.

\section{DFTB Allows for the Study of Hierarchical Nanostructures}

Nanoparticles rarely exist in isolation. Advances in lithographic and self-assembly techniques have enabled their arrangement into hierarchical structures with nanometer-scale precision.\cite{Mirkin1996,Girard2019, Mann2021, Calcaterra_2024, Chellam_2025_2} Nanoparticle aggregates and/or superlattices offer additional tunability of their optical properties, wherein the extremely large localized electric fields between plasmonic nanoparticles at small gap sizes enable optical nonlinearities and additional effects extending into the quantum regime.\cite{Shen2017,Zhang2022,Prodan2003-2,Bauman2022,Selenius2017} Classically, as the distance between two nanoparticles is reduced, the dipolar plasmon resonance is found to continuously redshift with a concomitant increase in their electric fields. However, once entering the extreme nearly-touching regime, the classical description of these gaps breaks down and electrons can tunnel between nanoparticles.\cite{McMahon2012,deNijs2017,Liu2023} From a quantum standpoint, the emergence of such charge transfer plasmons in the near-to-mid IR results in significantly reduced hybridization effects and damping of the electromagnetic field across the junction.\cite{Wen2015, Wu_2013, Zhu_2016} Accordingly, these effects cannot be fully explained using higher-level classical methods; one may imagine a nonlocal description required for separations below 5 nm and a fully quantum model at even smaller gap sizes.\cite{Garcia2008,McMahon2012, McMahon_2010, Wu2024}

\begin{figure}[h]
    \centering
    \includegraphics[width=0.8\linewidth]{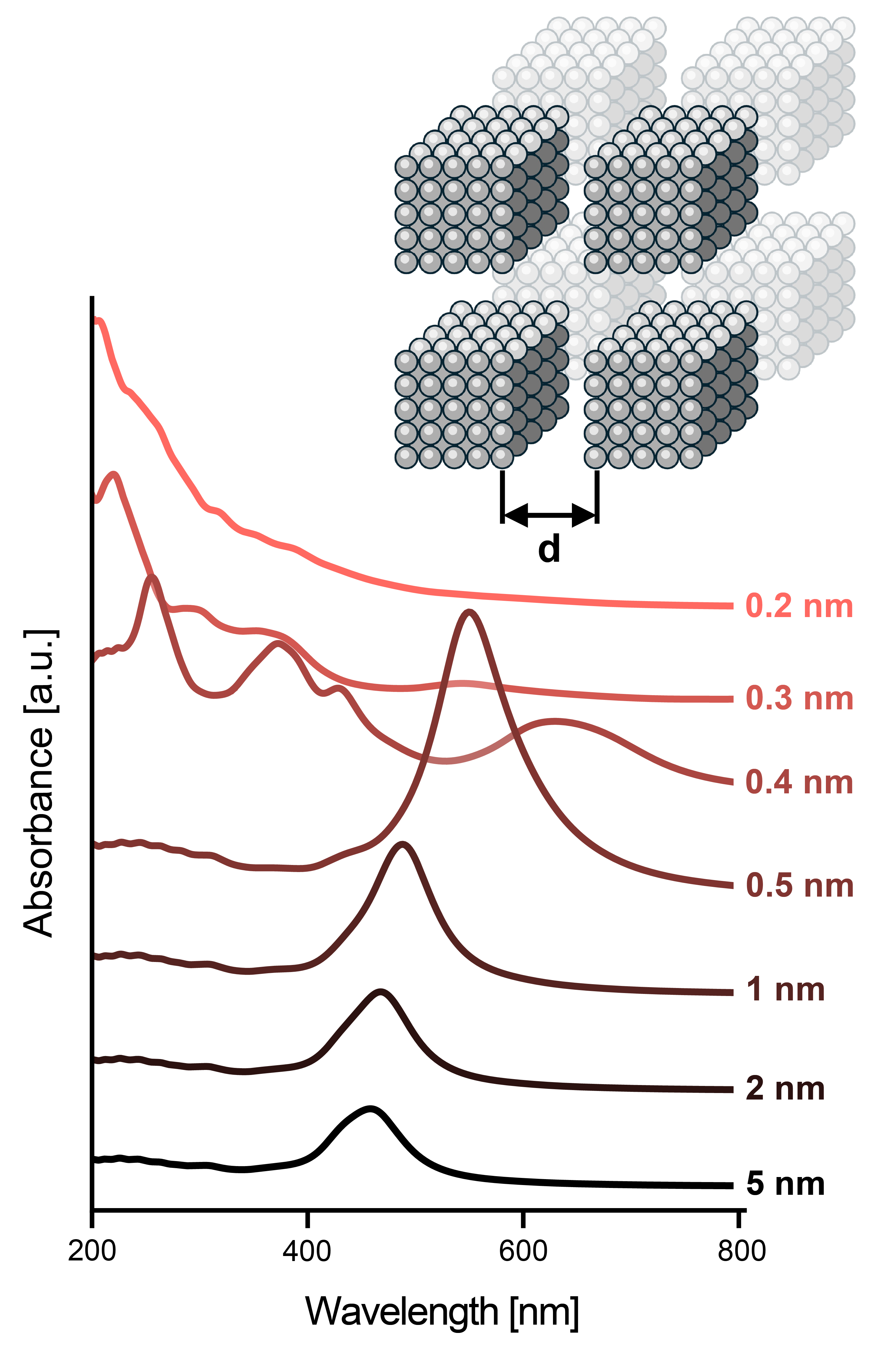}
    \caption{DFTB captures plasmon coupling in nanoparticle aggregates. Schematic of periodic boundary conditions consisting of \ch{Ag666} nanocubes separated by an interparticle spacing $d$. Corresponding absorption spectra with varying $d$ between 0.2 and 5 nm. The density matrix was propagated for 100 fs, with calculations performed using the {\asciifamily hyb-0-2} parameter set.}
    \label{fig:aggregate}
\end{figure}

We consequently simulated the absorption spectra for \ch{Ag666} nanocubes while employing periodic boundary conditions at varying gap distance $d$. From \textbf{Fig. \ref{fig:aggregate}}, we can see that as $d$ decreases from 5 nm to 0.5 nm, the plasmon wavelength redshifts and, concomitantly, the oscillator strength significantly increases, consistent with well-known plasmon coupling between adjacent particles. However, as $d$ further decreases below 0.5 nm, tunneling effects quench the plasmon response to the point where it is effectively unobservable at 0.2 nm. At 0.4 nm, several new modes are found; however, these do not show variation with decreasing $d$ that has been observed experimentally or with electrodynamic models.\cite{Wen2015} Thus, in contrast to a previous jellium model, which describes this classical-to-quantum onset at 1 nm,\cite{Zuloaga2009} DFTB shows this onset below 0.5 nm. The presence of charge transfer plasmons has been theorized to emerge in icosahedral \ch{Ag66} chains below 0.1 nm at the DFTB level, though at 2.5 eV -- a much higher energy than is typically expected (\textit{cf.} Ref. \citenum{Ilawe_2018}). Note also that a monotonic increase in gap distance for this transition was experimentally found with particle size, ranging from 0.8 to 1 nm for 30 to 80 nm nanoparticle dimers.\cite{Jose2022} 

Similar to the limitations discussed for alkali metal clusters, while DFTB is capable of simulating ground state electron transport,\cite{Hourahine2020, Pecchia_2004} it is thus far incapable of fully describing long-range excited state charge transport through these sub-10 nm junctions. The lack thereof consequently underestimates the interparticle distance threshold between classical and quantum effects and fails to properly describe charge transfer plasmons altogether. An extension of screened range-separated hybrid functionals has been specifically implemented for periodic systems in the context of KS-DFT and could mitigate this discrepancy;\cite{vanderHeide_2023} implementation thereof to DFTB would nonetheless require a separate parameterization for the (un)screened long-range Coulomb kernel. Small separation distances necessitate a quantum mechanical description to predict reliable electric field enhancements, while long-range Coulombic effects are necessary to fully describe coupling and charge transfer in nanoparticle aggregates.

Interparticle gaps also enable strong optical nonlinearities in the near-touching regime. As Ehrenfest dynamics is not limited to linear response theory as is the case for LR-TDDFT(B), the use of RT-TDDFTB (per Eq. 12) allows for the investigation of nonlinear optical phenomena. Plasmon resonant effects can amplify second (SHG) and third harmonic generation (THG) by several orders of magnitude due to the plasmons' ability to enhance the electromagnetic field in the vicinity of the particle. Our group has explored the use of RT-TDDFTB to study infrared upconversion in nanorod dimers.\cite{Giri-Schatz-2024-Plasmon-3} Electron dynamics were initiated \textit{via} a Gaussian envelope pulse. As these nanoparticles have a high density of electronic states near the plasmon resonance, we hypothesized that the use of a Gaussian driving profile (which features a more gradual increase in the field amplitude concomitant with excited state dephasing) would allow for more efficient harmonic generation (\textit{cf.} the linear ramp over one optical period employed in Ref. \citenum{Li_2013}). Harmonic generation signals were calculated from the induced dipole moment $\mu_{a}(t) = \text{Tr} \left[ (\hat{e}_{a} \cdot \hat{\mu}) \rho(t) \right]$ for Cartesian coordinate a, which coherently oscillates without bound in the absence of damping. Excitation thereof includes both fundamental and generated harmonics alongside all other activated modes:
\begin{align}
    \mu_{a}(t) &{}= \mu_{a}^{0} + \sum_{b}\int_{-\infty}^{\infty}\alpha_{ab}(t-t_{1})E_{b}(t_1)\,dt_{1}\\
    &+ \frac{1}{2!}\sum_{bc}\iint_{-\infty}^{\infty}\beta_{abc}(t-t_{1},t-t_{2})E_{b}(t_1)E_{c}(t_2)\,dt_{1}\,dt_{2}\nonumber\\
    &+ \frac{1}{3!}\sum_{bcd}\iiint_{-\infty}^{\infty}\gamma_{abcd}(t-t_1,t-t_2,t-t_3)\nonumber\\
    &\times E_{b}(t_1)E_c(t_2)E_d(t_3)\,dt_1\,dt_2\,dt_3+...,\nonumber
\end{align}
where $a$, $b$, $c$, and $d$ denote Cartesian coordinates, $\alpha$ is the polarizability, $\beta$, and $\gamma$ are the first and second hyperpolarizability tensors, respectively. Additionally, $\mu_{a}^{0}$ is the permanent dipole moment. 

\begin{figure}[h!]
    \centering
    \includegraphics[width = 0.9\linewidth]{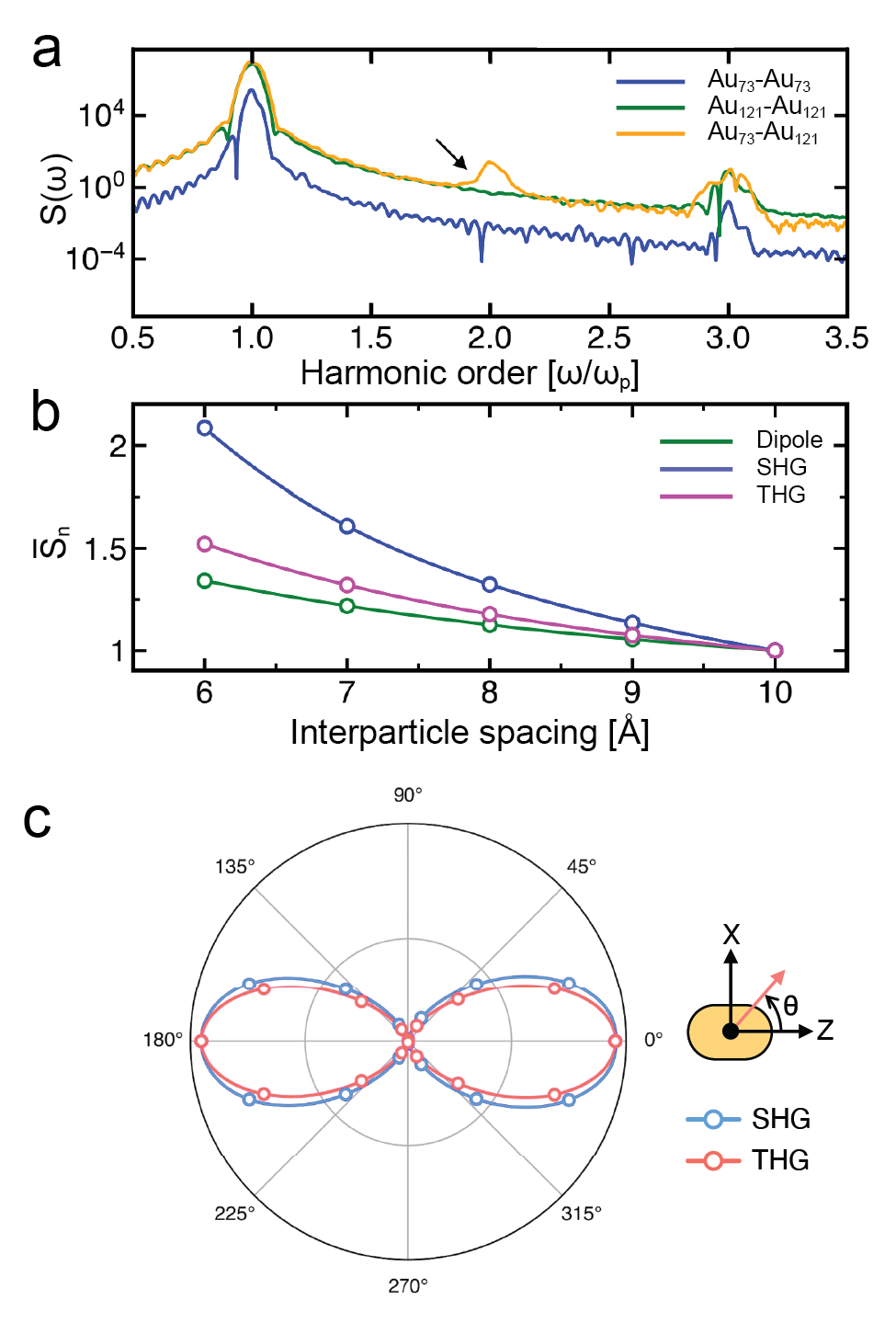}
    \caption{Asymmetric nanorod dimers arranged in end-to-end configurations allow for harmonic generation. \textbf{(a)} Harmonic generation for symmetric \ch{Au73-Au73} and \ch{Au121-Au121} dimers and an asymmetric \ch{Au73-Au121} dimer. Here, the interparticle spacing is fixed at 0.6 nm. \textbf{(b)} Dipolar, and S/THG signals as a function of interparticle spacing. The driving field intensity is fixed at $I = 2.5\times 10^{8}$ W/cm$^{2}$. All dimers were irradiated with a driving frequency at the plasmon frequency $\omega_p$ = 1.45 eV, and all spectra are normalized to the maximum peak intensity. \textbf{(c)} Integrated SHG and THG signals with respect to pump polarization for \ch{Au73-Au97}, making an angle $\theta$ with the $Z$ axis. The dimer was irradiated with an intensity of $I_{p} = 9.7\times 10^{7}$ W/cm$^{2}$ and the peak amplitudes at $\theta = 0^{\circ}$ are normalized to 1. Reprinted with permission from Ref.\citenum{Giri-Schatz-2024-Plasmon-3}. \textcopyright\, 2024 American Institute of Physics.}
   \label{polar_fig}
\end{figure}

The resulting HG signals exhibit characteristics that are consistent with previous TDDFT-based results. Modulating the particle length and interparticle spacing controls the longitudinal plasmon energy between 1.2 and 1.6 eV and, when excited, allows for S/THG. We specifically used end-to-end dimer configurations as particle symmetry breaking is a requisite condition for these processes: detectable SHG signals are observed only in nanorod dimer systems with broken longitudinal symmetry (note the differing spectra between the a/symmetric \ch{Au73} and \ch{Au121} dimers in \textbf{Fig. \ref{polar_fig}a}). The HG yield increases with particle size, and smaller interparticle spacings are found to enhance these processes. The generated signal intensity follows a power law dependence with respect to the incident electric field intensity, with exponents comparable to those determined \textit{via} perturbation theory \textbf{(Fig. \ref{polar_fig}b)}. Importantly, the largest system size studied in this work consists of 218 Au atoms, which is the largest nanoparticle system treated quantum mechanically in the context of harmonic generation thus far.

The maximum HG yield is observed when the laser polarization is aligned with the nanorods' longitudinal axes (\textbf{Fig. \ref{polar_fig}c}). Both the S/THG amplitudes tend to zero as the polarization angle deviates from longitudinal alignment, indicating that resonant excitation of the plasmon mode facilitates efficient harmonic generation. Similar trends in the polarization dependence of SHG have been previously reported in nanorod systems and nanosphere dimer systems.\cite{Maekawa_2020, Li_2021} Nevertheless, the THG amplitude decreases more rapidly compared to the SHG peak due to its higher sensitivity to the laser pulse.

From these spectra, we further calculated the polarizabilities using Eq. \ref{eq_pol} and hyperpolarizabilities through finite difference calculations of the dipole amplitudes and subsequent dipole fitting.\cite{Li_2013} In Table \ref{table_pol}, we list the XX, YY, and ZZ components of $\alpha$, the ZZZ component of $\beta$, and the ZZZZ component of $\gamma$. For higher-order signals, we report only the Z components as longitudinal excitation maximizes the HG yield while transverse excitation has a negligible effect (from \textbf{Fig. \ref{polar_fig}c}). Notably, the polarizabilities obtained from RT-TDDFTB agree with those obtained from LR-TDDFTB. The ZZ, ZZZ, and ZZZZ components are shown at their respective longitudinal plasmon energies, while the XX and YY components of $\alpha$ are presented at the transverse plasmon energy of 2.05 eV.

\begin{table}[h]
\centering
\begin{threeparttable}
    \setlength{\tabcolsep}{0.1cm}
    \renewcommand{\arraystretch}{1.4}
    \begin{tabular}{|c|c|c|c|c|c|c|}
        \hline
         \multirow{2}{*}{Dimer} & \multicolumn{3}{c|}{$\alpha$ [cm$^3$]} & $\beta_{zzz}$ & $\gamma_{zzzz}$\\
         \cline{2-4}
    & $\alpha_{xx}$ & $\alpha_{yy}$ & $\alpha_{zz}$ & [10$^{-26}$ esu]& [$10^{10}$ a.u.] \\
    & \big[10$^{-22}$\big] & \big[10$^{-22}$\big] & \big[10$^{-20}$\big] &  &  \\
         \hline
         \hline
         \ch{Au73-Au97} &1.29&1.29&2.34& 2.1 & - \\
         \ch{Au73-Au121} &1.47&1.47&4.89& 2.2 & 8 \\ 
         \ch{Au73-Au133} &1.51&1.51&5.94& 3.2 & 141.3 \\ 
         \ch{Au73-Au145} &1.75&1.75&9.76& 3 & 256.5 \\ 
         \hline
    \end{tabular}
  \end{threeparttable}
    \caption{Imaginary component of the polarizability $\alpha$, first hyperpolarizability $\beta$, and second hyperpolarizability $\gamma$. These values were evaluated using the {\asciifamily auorg-1-1} Slater-Koster parameter set. Results were evaluated at the longitudinal plasmon energies for $\alpha_{ZZ}$, $\beta_{ZZZ}$, and $\gamma_{ZZZZ}$. The remaining $\alpha$ components (XX and YY) were determined at the transverse plasmon excitation energy. Reprinted with permission from Ref.\citenum{Giri-Schatz-2024-Plasmon-3}. \textcopyright\, 2024 American Institute of Physics.}
    \label{table_pol}
\end{table}

All (hyper)polarizabilities monotonically increase with system size. These findings align with previously reported linear absorption calculations for icosahedral Ag and Au nanoparticles using TDDFTB showing absorption cross sections on the order of $10^{-15}$ to $10^{-14}$ cm$^2$.\cite{Douglas-Gallardo2019} Moreover, the first order hyperpolarizabilities are comparable to those experimentally obtained by Ngo \textit{et al.}, with $\beta \sim 10^{-26} - 10^{-25}$ esu for nanosphere and nanorod dimers with particle diameters as small as 3 nm.\cite{Ngo_2016} As is to be expected, these values are significantly greater than static hyperpolarizabilities obtained \textit{via} TDDFT for Au clusters, with reported values $\sim 10^{-29}$ esu.\cite{Pei_2022} While several THG measurements of Au nanoparticle films have been documented,\cite{Obermeier_2018} these studies do not report second-order hyperpolarizabilities, thereby limiting a direct comparison. Our obtained values for $\gamma \sim 10^{12}$ a.u. (or $5 \times 10^{-28}$ esu) are about 1000 times greater than the highest values reported for organic molecules.\cite{Tykwinski_1998} We consider this a resonable difference given our focus on resonant responses, as opposed to the nonresonant responses studied therein.

Linear response theory, as its name suggests, is only capable of simulating linear absorption spectra. While it may suffice for small, isolated nanoparticles, higher-order structures present a more complex picture. To this end, real-time dynamics, in tandem with the computational tractability of DFTB, offers the ability to model nonlinear optical phenomena with reasonable accuracy. While the distance by which tunneling effects are greatly underestimated due to the short-ranged interactions prevalent within this construction, RT-TDDFTB allows for a detailed analysis of the nanoparticle gap region, which contains highly confined electromagnetic fields and consequent optical nonlinearities and nonlocalities. Similar limitations when considering the optical properties of individual nanoparticles present in the context of their aggregates. The incorporation of longer-ranged Coulombic interactions, much in the same way as has been done for organic molecules,\cite{Vuong_2018, Darghouth_2021} could enable more accurate comparisons to experiment. Beyond studying nonlinearities in the small gap regime, refining these quantum methods could enable a systematic analysis of cavity-mediated chemistry.\cite{Hirai2023,Mandal2023,Bhuyan2023} More broadly, semiempirical methods such as DFTB allow for scalable, parameter-free studies of real-time polaritonic chemistry in nanoparticle aggregates and dimers, in contrast to prior studies limited by system size.\cite{Reits_2025, Rossi2019,Kuisma2022,Schafer2022} Doing so would allow for a better understanding of the rich physics underlying these processes beyond experimentation alone.

\section{DFTB Enables Tractable Studies of Plasmon-driven Photocatalysis}
Plasmons ultimately occur on the femtosecond timescale. Following plasmon excitation, plasmons can either decay radiatively (a slower process involving photon re-emission\cite{Ross2015}), or nonradiatively by dephasing to electron-hole pairs with energies equivalent to that of the plasmon.\cite{Brongersma2015, Stefancu_2024} These energetically ``hot" carriers subsequently redistribute their energy through scattering events with photons and other low-energy electrons over 100 fs to 1 ps. As these scattered electrons continue to lose energy, their propensity to interact with phonons increases between 100 ps to 10 ns, leading to thermal dissipation into the surrounding environment. Together, these processes can raise the electrons' translational temperatures to thousands of degrees higher than their surroundings (depending on nanoparticle size), creating an efficient mechanism for electron transfer into nearby molecules that bypasses scaling relationships that would otherwise limit catalytic turnover. Importantly, plasmon-driven catalysis can offer milder reaction conditions while directly harnessing light to drive such chemical transformations.

While the hot carrier picture remains the prevailing theory governing plasmon-mediated photocatalysis, other mechanisms have nonetheless been proposed. Direct electronic excitation between metal-adsorbate states,\cite{Kazuma_2020, Stefancu_2024} plasmon electric field enhancement of adsorbate photochemistry\cite{Seemala_2019}, and enhancement by photothermal heating\cite{Zhou2018} have all been hypothesized to be capable of driving many reactions of interest. Difficulties in observing plasmon dynamics at inherently short length and time scales, however, bring about questions relating their dynamics to any observed catalytic effects, leaving some mechanistic questions best answered \emph{via} theoretical approaches.\cite{Christopher2011, Zhou2018, Li2023, Hartland2017, Cortes2020, Kumar2019} In this regard, our group has focused on the theoretical analysis of plasmon-driven photocatalysis, whereby disentangling the many events underlying these processes should allow for a complete description of hot electron dynamics and their intimate role in governing the outcomes of chemical reactions. We have previously investigated the mechanism for photoinduced \ch{H2} dissociation on Au nanoclusters using TDDFT.\cite{Wu2020} The dissociation of \ch{H2} is found to be governed by transitions from metallic ``hot electron" states to charge transfer states \emph{via} the antibonding $\sigma^{*}$ orbital on \ch{H2}. However, this comprehensive study was limited to studying \ch{H2} near an \ch{Au6} cluster -- while this cluster showed a ``plasmon-like" peak in its TDDFT absorption spectra, it was too small to support true plasmons and therefore could not accurately model the plasmon dephasing processes that would otherwise generate hot electrons.

It then follows that first-principles modeling of these processes is prohibitively challenging due to the large number of electrons and excited states involved. This is particularly the case for metals (and especially so for free-electron metals) where their continuous density of states makes electronic convergence difficult, if not impossible, for systems large enough to exhibit plasmonic behavior.\cite{Seveur2023} Moreover, the tens to hundreds of fs timescales involved in plasmon-driven photodissociation force a compromise in electronic structure studies that leads to the consideration of small clusters and short pulses (below ten fs).\cite{Yan_2015, Zhang2018, Kuda_2021, Chen_2023, Li2023} These small clusters serve as surrogate models for investigating photocatalysis and require a high intensity threshold for any reaction to occur. Not only do these high intensities deviate from actual experimental conditions, but the short pulses and powers used therein (as in Ref. \citenum{Zhang2018}) approach or even exceed the limit where a relativistic description (\textit{e.g.} inverse bremsstrahlung\cite{Seely_1974}) is required.\cite{Takabe_2020}

\begin{figure}[h]
    \centering
    \includegraphics[width=0.7\linewidth]{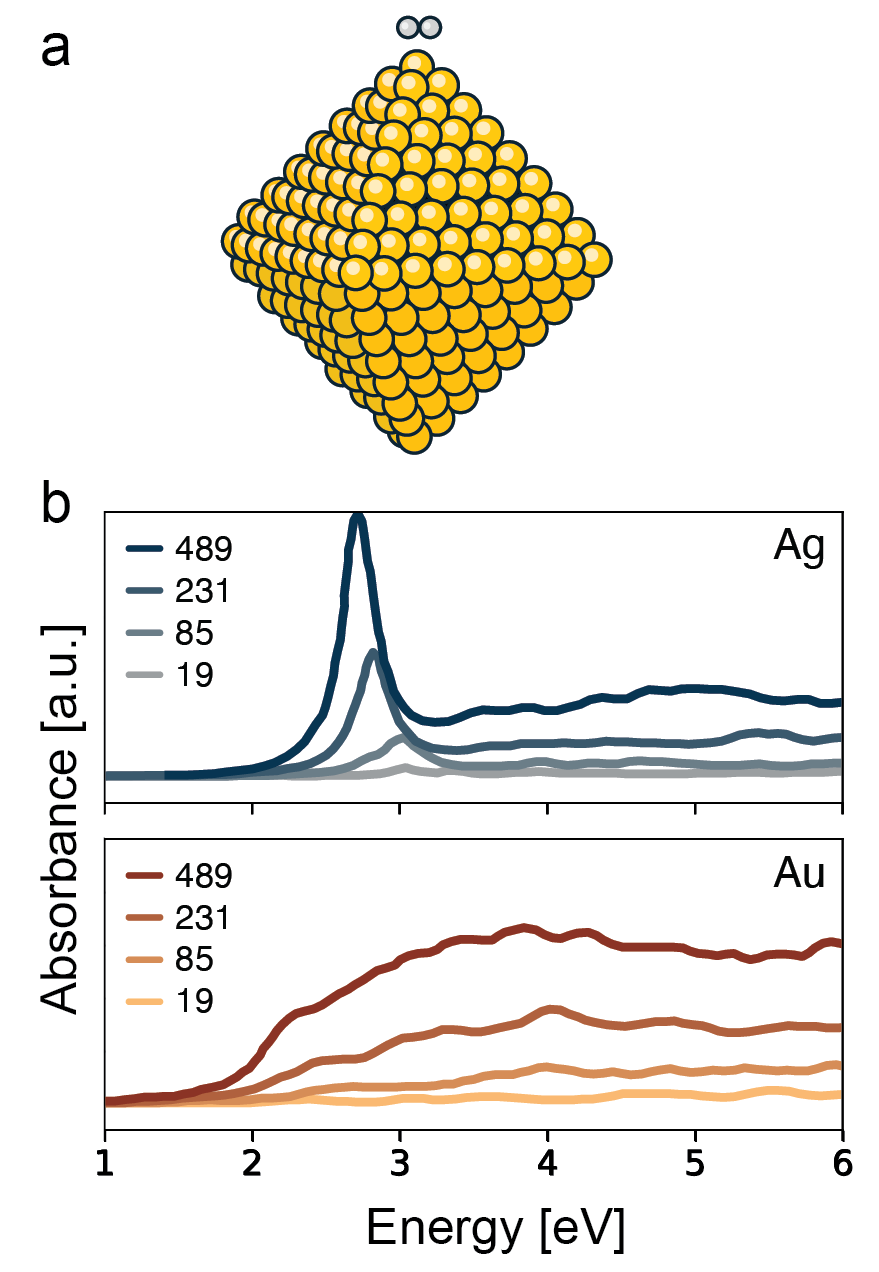}
    \caption{A \ch{H2} molecule adsorbed on octahedral Ag and Au nanoparticles of varying sizes. \textbf{(a)} Representative configuration of a \ch{Au489} nanoparticle with an \ch{H2} molecule positioned at the particle tip. \textbf{(b)} Corresponding absorption spectra for Ag and Au nanoparticles spanning 19 to 489 atoms. \ch{H2} adsorption minimally changes their absorption spectra. All geometries were optimized using the {\asciifamily hyb-0-2} and {\asciifamily auorg-1-1} Slater-Koster parameter sets for Ag and Au, respectively. The density matrix was propagated for 200 fs with a step size of $2\times 10^{-3}$ fs. Adapted with permission from Ref.\citenum{Giri2023}. \textcopyright\, 2023 American Chemical Society.}
    \label{fig4}
\end{figure}

To this end, RT-TDDFTB provides a natural choice to model the reaction dynamics underpinning plasmon-mediated photocatalysis. The incorporation of derivative coupling in Eq. \ref{eq32} leads rather naturally to adsorbate dissociation, and, coupled with DFTB's scalability, allows for a systematic study of these processes at significantly larger sizes and longer time scales. We have examined the photodissociation of \ch{H2} on octahedral Au and Ag clusters as a function of particle size and driving frequency (\textit{i.e.}, plasmon \textit{vs.} interband) on the dissociation threshold intensity.\cite{Giri2023} Particles spanning 19 to 489 atoms (between 0.5 to 2.2 nm) were constructed with a \ch{H2} molecule positioned at the particle vertex (\textbf{Fig. \ref{fig4}a}). While different adsorption sites provide distinct chemical environments, we deliberately positioned the \ch{H2} molecule at the particle tip as this site experiences the maximum charge localization (and electric field enhancement) through plasmon excitation.

Our RT-TDDFTB-calculated absorption spectra align well with previous TDDFT studies,\cite{aikens2008, Gieseking2020} confirming the particles' plasmonic behavior arising from collective electronic excitation (\textbf{Fig. \ref{fig4}b}). For both Au and Ag nanoparticles, larger sizes run concomitant with stronger oscillator strengths, reflecting their increased density of states. The spectra reveal distinct characteristics for each metal. Ag particles exhibit a prominent $sp$ intraband transition (\textit{i.e.}, plasmon peak) between 2.5 and 3 eV, with additional $d \rightarrow sp$ interband transitions appearing as relatively uniform signals at higher energies. In contrast, Au shows a broader plasmon peak near 2.3 eV attenuated by overlapping $d\,\rightarrow sp$ interband transitions, making it difficult to distinguish plasmon excitation in smaller particles.\cite{Aikens2006} Only for Au$_{489}$ does the plasmon evolve into a spectrally distinct feature around 2.3 eV, reminiscent of that found in larger, spherical Au nanoparticles. Both metals demonstrate a consistent $\approx 0.25$ eV redshift as the particle size increases from 19 to 489 atoms. While this shift parallels previous observations in 0.5 to 2.0 nm tetrahedral Ag clusters,\cite{aikens2008} our results suggest quantum size effects and $d$ electron screening dictate this behavior, rather than the dynamic depolarization effects seen in much larger systems (\textit{vide supra}).

\begin{figure}[h]
    \centering
    \includegraphics[width = \linewidth]{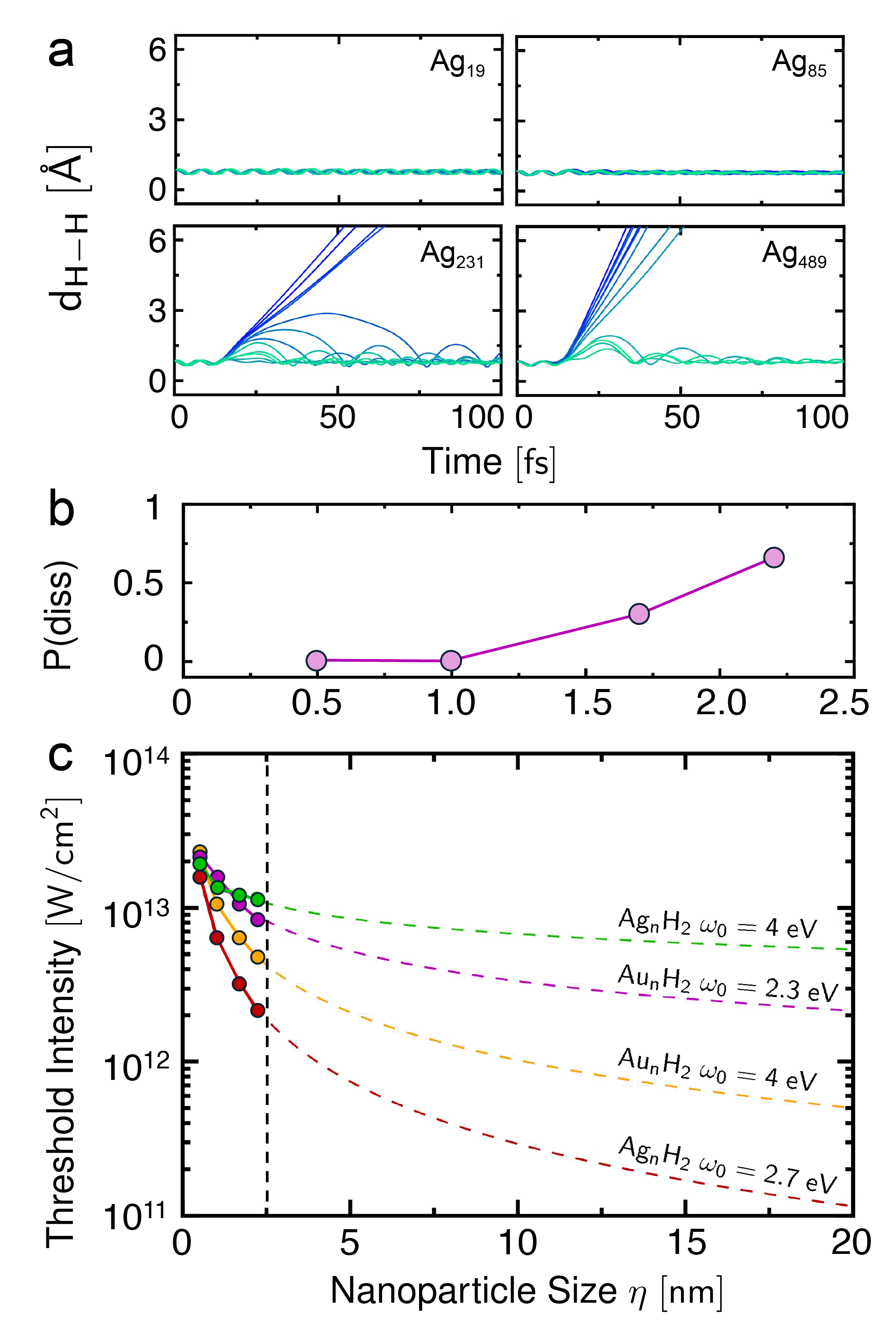}
    \caption{\ch{H2} dissociation probability increases with particle size. \textbf{(a)} H-H bond dissociation trajectories with increasing Ag nanoparticle size. A fixed laser peak intensity of $I=3.3\times10^{12}$ W/cm$^2$ (threshold intensity for \ch{Ag231}) is used. \textbf{(b)} Dissociation probabilities with particle size calculated at 100 fs. Particles were irradiated with a driving energy of 2.7 eV. \textbf{(c)} Decay of threshold intensity ($I_t$) with particle size ($\eta$) for the Ag$_n$H$_2$ and Au$_n$H$_2$ complexes, subject to an external driving with a frequency tuned to induce intra- and interband transitions (listed on the right of each curve). Adapted with permission from Ref.\citenum{Giri2023}. Copyright 2023 American Chemical Society.}
   \label{h2_disso_fig}
\end{figure}

With the particles' spectra fully resolved, we hypothesized that RT-TDDFTB could be used to investigate the relative efficacies of plasmon \emph{vs.} interband excitation for \ch{H2} dissociation. Doing so would enable a statistically meaningful study correlating particle size, driving frequency, and consequent catalytic turnover. To do so, a Gaussian laser pulse was used to excite the nanoparticle-molecule complex. Modulating the driving frequency can therefore probe the comparative dynamics between inter- and intraband transitions. To do so, we simulated a Gaussian pulsed excitation centered around 25 fs at the corresponding driving frequencies of $\omega_0 = 2.7$ eV (2.3 eV) for the Ag (Au) plasmon, and 4 eV at the interband transition. The pulse amplitude decays to zero after approximately 25 fs, resulting in a broad energy band centered around its central frequency, rather than a delta function energy as experimentally seen with continuous-wave lasers.

The probability of \ch{H2} dissociation strongly depends on nanoparticle size. To investigate this relationship, we simulated 20 trajectories for Ag nanoparticles ranging from \ch{Ag19} to \ch{Ag489} at a fixed laser intensity of $3.3 \times 10^{12}$ W/cm$^{2}$. Initial electronic velocities were sampled from a Maxwell-Boltzmann distribution at 300 K to allow for a nondeterministic outcome \textbf{(Fig. \ref{h2_disso_fig}a)}. For smaller particles at this sub-threshold intensity, \ch{H2} molecules remained intact, showing only coherent oscillations of the the H-H bond distance ($d_{\rm H-H}$) around its equilibrium value. Larger particles, however, show an increased dissociation probability, with some trajectories achieving \ch{H2} dissociation within 25 fs \textbf{(Fig. \ref{h2_disso_fig}b)}. Concurrently, some trajectories exhibited strong $d_{\rm H-H}$ oscillations between 1.5-2 \AA\, before damping to values similar to those seen for smaller particles. This size-dependent behavior can be explained by the inverse relationship between particle size and its dissociation threshold -- that is, the minimum intensity at which \ch{H2} dissociation occurred. Larger particles can sustain stronger dipoles upon excitation, leading to more efficient hot electron generation and transfer to adsorbates.

Nonetheless, extrapolation to larger sizes (from the 2.5 nm used herein to 20 nm as used experimentally\cite{Keller_2018}; \textbf{Fig. \ref{h2_disso_fig}c}) reveals a disparity in the dissociation intensity by three orders of magnitude (\textit{cf.} the threshold intensity for free \ch{H2} in Ref. \citenum{Ludwig_1997}). Several limitations in our model may reconcile this discrepancy.\cite{Kar_2025} For one, the use of longer pulse durations (as used experimentally, $\sim 10^2$ fs) would mitigate electronic relaxation due to limitations in describing electron-electron scattering at this level of theory; Ehrenfest dynamics do not fully capture stochastic energy dissipation due to an incomplete many-body treatment. For another, a significant portion of the excitation energy is redistributed towards other processes instead of directly driving \ch{H2} dissociation. As only one \ch{H2} molecule is considered per particle, plasmon excitation does not lead to efficient energy transfer in this model. In reality, multiple molecules would be in proximity to the particle (alongside particle aggregates with higher localized electromagnetic fields), where plasmonic hot carriers could interact collectively with the reactant ensemble and enhance the dissociation efficiency.\cite{Kar_2025} Similar approaches to those examined in the previous section could be used to understand the reaction efficiency in idealized nanoparticle aggregates, while nonetheless raising the complexity of the problem.

\begin{figure}[h]
    \centering
    \includegraphics[width = \linewidth]{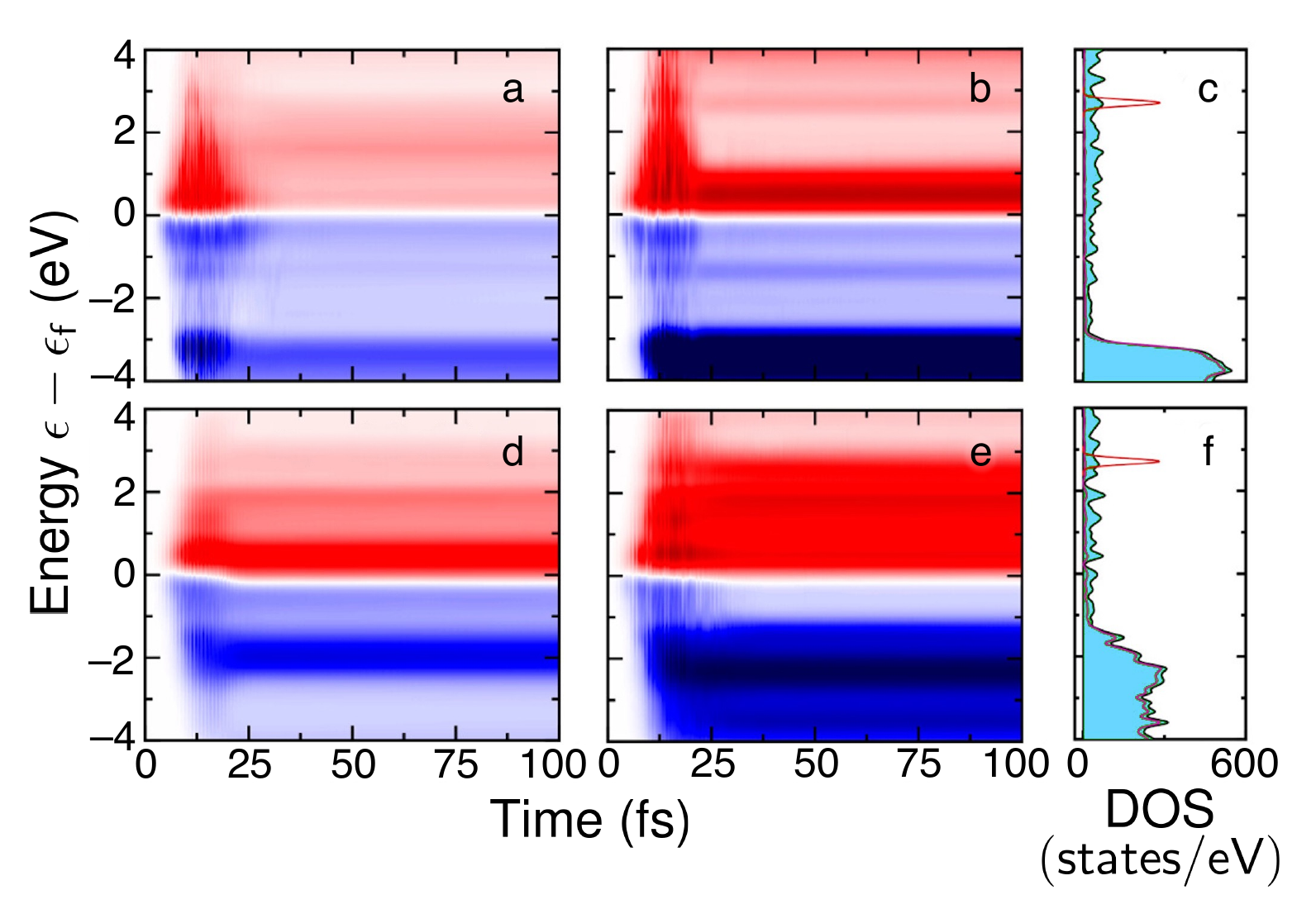}
    \caption{Orbital populations as a function of energy and time for driving frequencies \textbf{(a)} 2.7 and \textbf{(b)} 4 eV for Ag and \textbf{(d)} 2.3 and \textbf{(e)} 4 eV for Au complexes. The red (blue) color represents the population gain (depopulation) with positive (negative) amplitudes. Density of states (DOS) where magenta lines represent the \textit{d} band projected DOS for \textbf{(c)} Ag and \textbf{(f)} Au complexes. The red lines in the DOS panels correspond to the \ch{H2} $\sigma^{*}$ antibonding orbital. Adapted with permission from Ref.\citenum{Giri2023}. Copyright 2023 American Chemical Society.}  
   \label{mopop_fig}
\end{figure}

In contrast to Ag, where the plasmon is found to be more efficient in driving \ch{H2} dissociation, interband excitation in Au is found to be more effective \textbf{(Fig. \ref{h2_disso_fig}c)}. Analysis of the electron population dynamics reveals key mechanistic differences between the two excitation modes. In every case, oscillations of the electron population in resonance with the laser frequency (known as ``sloshing" and ``inversion") initially appear and quickly dephase after the pulse ends. During the pulse, sufficient populations accumulate near the \ch{H2} $\sigma^{*}$ orbital (represented by the red lines in \textbf{Fig. \ref{mopop_fig}c} and \textbf{f}). For both Ag and Au, plasmon excitation predominately populates states near the Fermi level, while interband excitation accesses higher-energy states through $d \rightarrow sp$ transitions. Au nanoparticles, however, exhibit a broader $d$ band energy range closer to the Fermi level compared to Ag, which enables significant interband transitions above the Fermi level. These higher-energy electrons transfer more effectively into \ch{H2}'s $\sigma^{*}$ orbital, weakening the molecular bond and inducing dissociation compared to lower-energy plasmon excitations which are further attenuated due to overlap with interband modes. However, a competing argument claims that population transfer is dictated by the complexity of its dynamics rather than through the spectral overlap between bands.\cite{Berdakin_2020} Specifically, through DFTB, plasmon excitation of Ag nanoparticles is found to continuously yield $sp$ electron-hole pairs with energies in proximity to the Fermi level, while Au initially produces low energy $sp$ holes that get depopulated in favor of high energy holes at the $d$ band threshold at longer times. 

Building on these studies, our recent work using RT-TDDFTB has explored plasmon-driven electrochemical selectivity between \ch{H2O} oxidation and \ch{CO2} reduction.\cite{Alcorn_2024} Experimentally, plasmon excitation of Au-Cu alloy nanoparticles was found to shift the reaction selectivity of aqueous \ch{CO2} electroreduction, favoring \ch{H2} evolution over \ch{CO} formation. Concomitantly, we found through RT-TDDFTB that hole-driven \ch{O-H} bond dissociation dominates over electron-driven \ch{C-O} bond-breaking in \ch{Au365} nanocubes. While Au-Cu Slater-Koster parameters have not yet been evaluated, the similar dielectric functions of Au and Cu suggest the applicability of this analysis to the experimental Au-Cu system.

Thus far, the use of RT-TDDFTB outright is limited to the first tens of femtoseconds when analyzing the electron dynamics associated with plasmon-driven photochemistry. While in principle DFTB's large computational tractability enables studies into the picosecond regime, inefficient descriptions of electronic relaxation pathways limit the means by which longer-time effects (\textit{e.g.} electron-electron/phonon scattering) can be examined in practice.\cite{Sanchez_2022} To examine these limitations, we have investigated the plasmon-induced electron dynamics in Al nanoparticles, where, unlike Au or Ag, Al's free-electron character allows for efficient electronic thermalization without interband coupling.\cite{Chellam_2025} Plasmon excitation and subsequent analysis of the electron dynamics for \ch{H2} dissociation found that multiphoton absorption led to the formation of distinct step-like features in the particles' transient electron distribution. The spacing thereof is commensurate with the photon energy ($\hbar\omega$) and reached energies up to 20 eV, which induced efficient \ch{H2} dissociation. However, electronic thermalization only occurred during the pulsed excitation. Contrary to semiclassical two-temperature models commonly used to describe electron dynamics\cite{Mueller_2013}, these step-like features persisted rather than dissipating into an equilibrium Fermi-Dirac distribution, highlighting the limitations of electronic relaxation modeling within the RT-TDDFTB framework. 

In other words, Ehrenfest dynamics' mean-field description of electron-phonon coupling and electron-electron scattering does not entirely describe the many relaxation channels involved in plasmon decay. Electron-electron scattering can be in principle implemented through the GW approximation. It is worth noting that the GW approximation has already been applied to DFTB to account for quasiparticle energy corrections.\cite{Niehaus_2005} This theory is nonetheless highly sensitive to the quality of the Coulomb potential used to describe interelectronic interactions; it is unclear whether the minimal basis set and approximations to the Coulomb potential in DFTB is sufficient to accurately describe screening at the GW level (high computational memory demands notwithstanding). Further, the nonadiabatic coupling matrix as presented in Eq. \ref{eq32} allows for nuclei to be treated semiclassically and electrons quantum mechanically. Including this term during plasmon irradiation is found to induce breathing-like oscillations within the first few femtoseconds.\cite{Bonafe2017} This observation, however, is ascribed to electronic excitation into antibonding states and is unrelated to electron-phonon scattering which occurs over a much longer timescale. 

Methods such as correlated electron-ion dynamics can recover electronic thermalization by augmenting the equation of motion with extra terms that describe nuclear fluctuations.\cite{Horsfield_2005, Lively_2021} These approaches nevertheless become rapidly intractable; high-order polynomial scaling with respect to both the electron and phonon dispersion limits their use for the larger particles that DFTB is able to study.\cite{Horsfield_2005} Additionally, a recent DFTB implementation of Boltzmann transport theory in the context of ground state electron transport has been shown to reasonably account for electron-phonon coupling;\cite{Croy_2023} a similar extension to real-time excited state dynamics could be valuable to account for plasmon decay processes. An attempt to reconcile these differences has been studied by the use of trajectory surface hopping to study the long-time electronic dynamics associated with CO adsorption on an \ch{Au20} tetrahedron, which combined TDDFTB with \textit{ab initio} molecular dynamics.\cite{Wu_2023} In this model, plasmon-generated hot carriers transfer back and forth between CO and the \ch{Au20} nanoparticle, thereby activating the \ch{C+O} stretching mode through nonadiabatic coupling (\textit{i.e.}, electron-phonon coupling) on the picosecond timescale. This more rigorous treatment of electronic energy redistribution demonstrated a 40\% efficiency for plasmon-mediated \ch{CO} activation through ensemble averaging. 

Taken together, the use of RT-TDDFTB allows for one to investigate photocatalytic processes at computationally tractable length scales into the nanometer regime and time scales into the picosecond domain. The use thereof presents a closer reconciliation between theory and experiment while under a fully quantum representation. However, while dissipation is putatively accounted for in Ehrenfest dynamics, an incomplete many-body description of the scattering processes involved with plasmon de-excitation precludes a thorough analysis of these phenomena past the first tens of femtoseconds thus far. 

\section{High Computational Tractability Requires Low Levels of Theory}
Looking deeper into the quantum realm, it becomes increasingly clear that Feynman's vision was simultaneously too modest and too ambitious. Today's desktop workstations offer processing speeds that supercomputers from just a decade ago could only dream of. Yet the unfavorable $\mathcal{O}(N^{3-6})$ scaling of high-level quantum chemistry methods remains an insurmountable barrier for a routine investigation of plasmonic systems outright.\cite{Goedecker_1999} Low-level quantum methods, and in particular DFTB, may ultimately reveal more about plasmonic phemonena than their more sophisticated counterparts higher up the theoretical hierarchy. DFTB offers a two to three orders of magnitude improvement in computational efficiency while providing a fair description of excited state properties. This approach has been postulated and has existed for nearly four decades, yet it remains surprisingly underutilized. 

It is still surprising that such a tight-binding method can provide a reasonable picture of delocalized excited state dynamics, and yet our work and those from many others in the field have shown that it can in fact do so. This is not to suggest that DFTB is without its limitations. Significant opportunities exist to enhance these methods, particularly when modeling the electron scattering and relaxation processes necessary to describe their long-time excited state dynamics. Similarly, refining the description of the long-ranged Coulomb potential could resolve persistent discrepancies between theory and experiment, especially in the context of alkali metal plasmonics, nanoparticle aggregates, and electron transfer processes. And perhaps most importantly, the development of additional Slater-Koster parameters for both ground and excited states would extend these capabilities to a broader range of materials, allowing for investigations into their emergent properties and resultant chemistry. This is all to say that there is indeed plenty of room at the bottom -- at the bottom of Jacob's ladder.\cite{Perdew_2001}

\section{Author Information}
\subsection{Author Contributions}
Conceptualization: N.S.C. and G.C.S.; Methodology: N.S.C., S.K.G, and G.C.S.; Validation: N.S.C., S.C., A.G., and G.C.S.; Data Curation and Visualization: N.S.C. and S.C.; Writing - Original Draft: N.S.C.; Writing - Review \& Editing: N.S.C., S.C., A.G., S.K.G., and G.C.S; Supervision: G.C.S. This manuscript was written through contributions of all authors. All authors have given approval to the final version of the manuscript.

\subsection{Biographies}

\begin{figure}[h!]
    \includegraphics[width=0.3\textwidth]{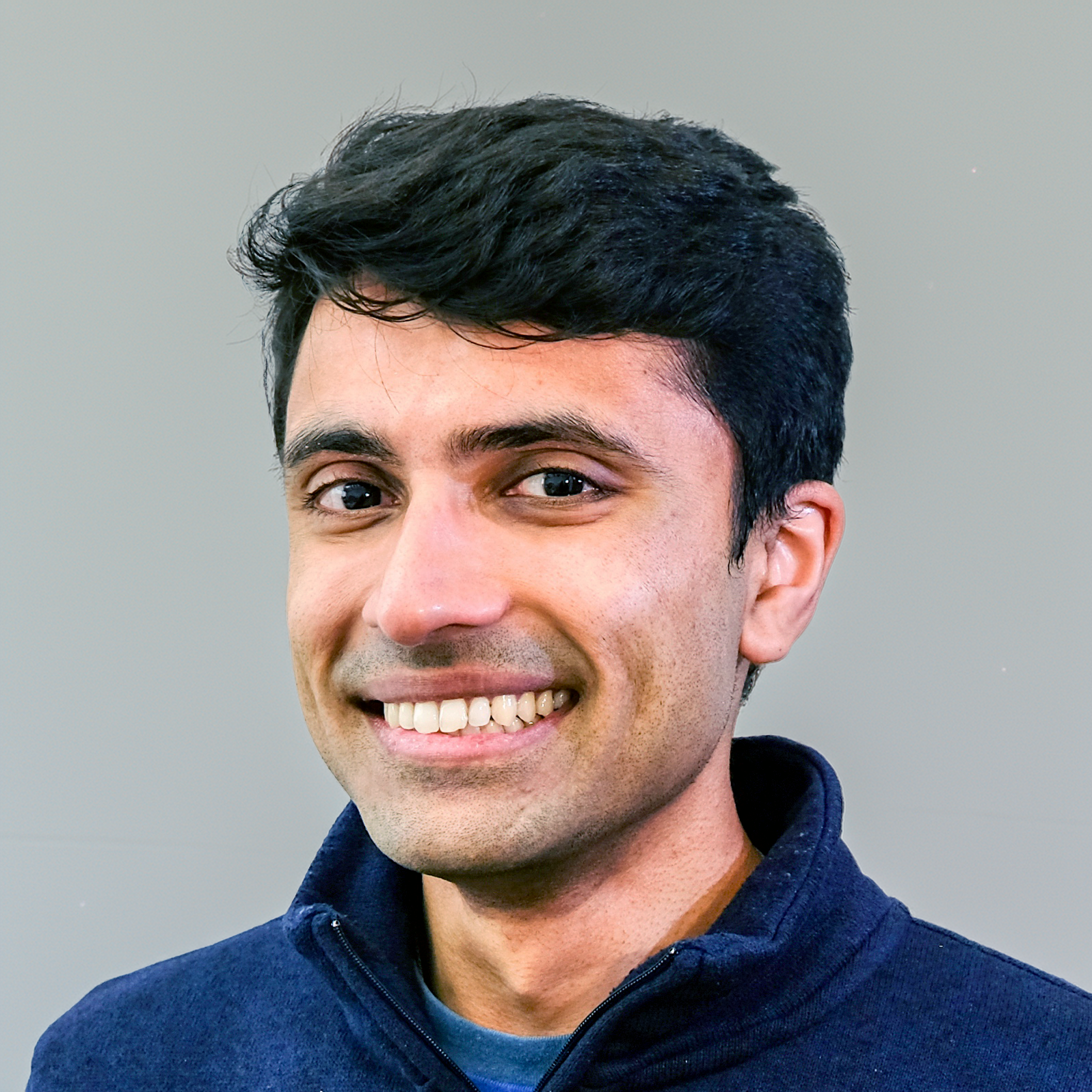}
\end{figure}

\noindent \textbf{Nikhil S. Chellam} received his Ph.D. in Chemical \& Biological Engineering from Northwestern University in March 2025, supported by a NSF Graduate Research Fellowship. Under the guidance of Profs. George C. Schatz and Chad A. Mirkin, his research focused on plasmons in aluminum nanocrystals alongside colloidal crystal engineering with DNA. Nikhil will begin his postdoctoral research training in July 2025 at MIT. His research interests broadly center on the fundamental physics and energy transport mechanisms governing catalytic systems. He earned a B.A. in History and a B.S.Ch.E. in Chemical \& Biomolecular Engineering from Rice University in 2020.

\begin{figure}[h!]
    \includegraphics[width=0.3\textwidth]{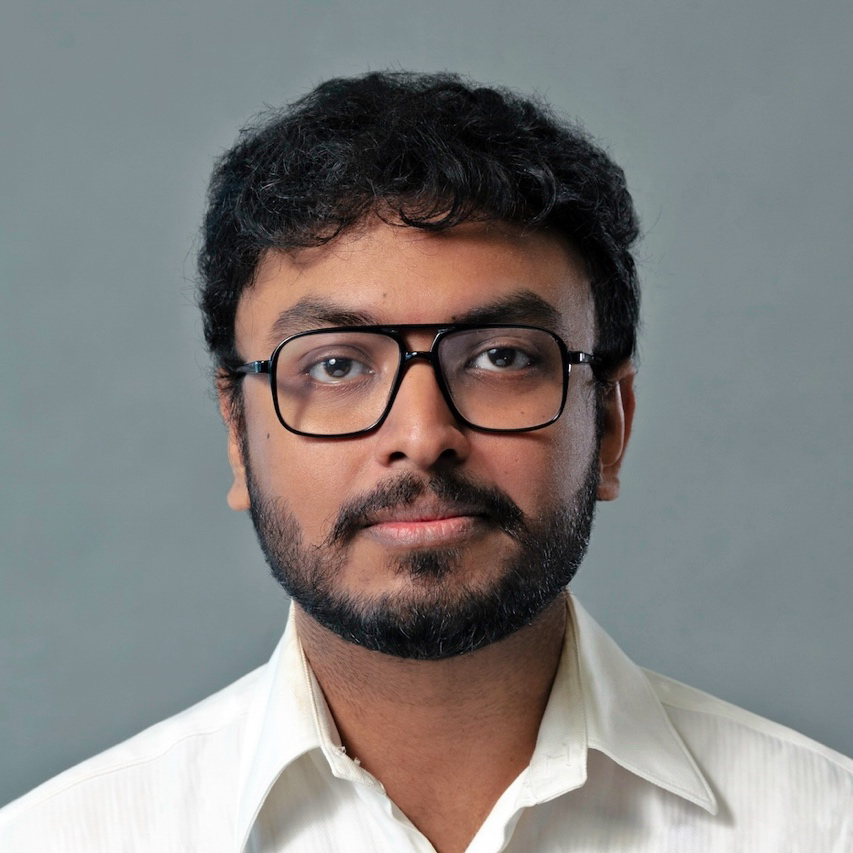}
\end{figure}

\noindent \textbf{Subhajyoti Chaudhuri} is an International Institute for Nanotechnology Postdoctoral Fellow working with Prof. George C. Schatz in the Department of Chemistry at Northwestern University. His research focuses on developing methods to study charge, spin, and energy transfer in complex systems. Subha received his Ph.D. in Applied Physics in 2020 from Yale University, where he developed methods to study electron transfer in weakly coupled systems with Prof. Victor S. Batista.

\begin{figure}[h!]
    \includegraphics[width=0.3\textwidth]{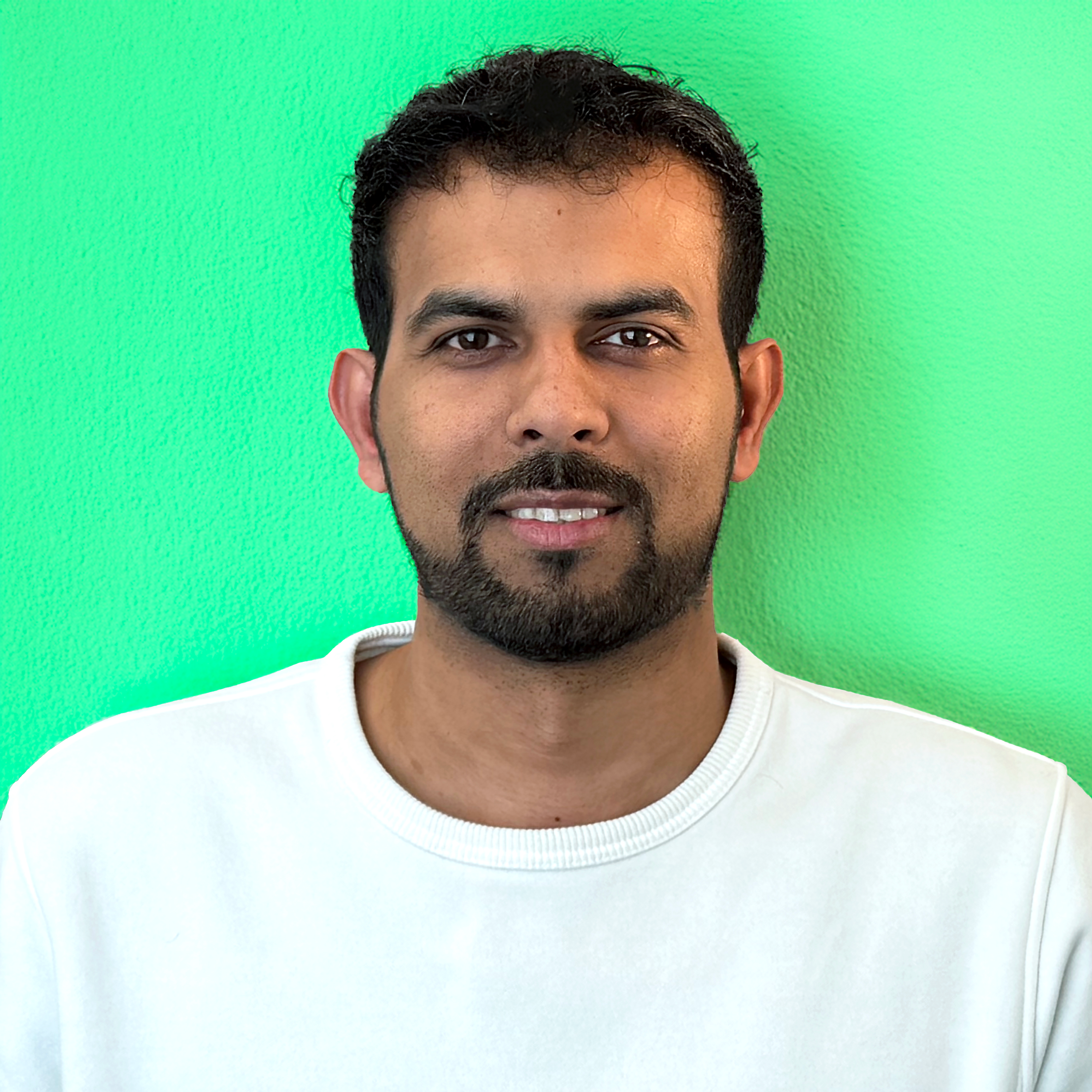}
\end{figure}

\noindent\textbf{Abhisek Ghosal} is a postdoctoral researcher in the Schatz Group at Northwestern University, where he works on advancing electronic structure theory and quantum dynamics. His current research addresses challenges in open quantum systems such as quantum plasmonics, charge transport in molecular junctions, and associative ionization in high-velocity atmospheric collisions. He received his Ph.D. in 2021 from the Indian Institute of Science Education and Research in Kolkata, where he earned the 2022 Molecular Physics Early Career Researcher Prize. Before coming to Northwestern, he was a postdoctoral fellow at the Tata Institute of Fundamental Research. He holds a B.Sc. and M.Sc. in Chemistry from Presidency College at the University of Calcutta.\\~\\

\begin{figure}[h!]
    \includegraphics[width=0.3\textwidth]{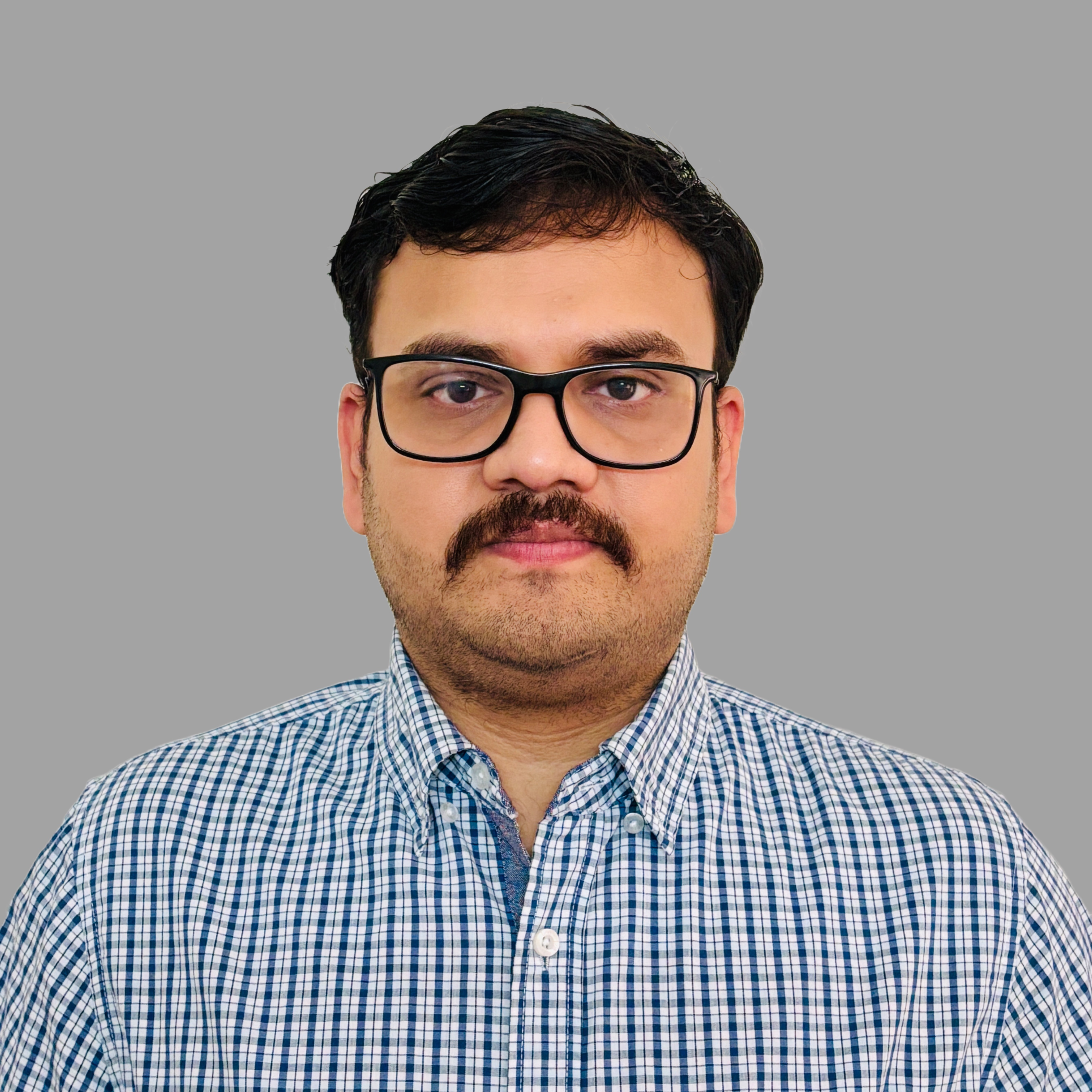}
\end{figure}

\noindent\textbf{Sajal Kumar Giri} is an Assistant Professor of Chemistry at IIT Dhanbad. Sajal obtained his Ph.D. in Chemical Physics from the Max Planck Institute for the Physics of Complex Systems under Prof. Jan Michael Rost. After industry research as a Machine Learning Scientist at Iambic Therapeutics (Entos) and postdoctoral research with Prof. George C. Schatz at Northwestern University, he joined IIT Dhanbad in 2024. His research interests focus on light-matter interactions with a particular emphasis on quantum optics, open quantum dynamics and machine learning methods to understand the quantum aspects of light-induced processes in molecular and nanoscale systems. Sajal holds a B.Sc. in Chemistry from Ramakrishna Mission Vidyamandira and an M.Sc. in Chemistry from IIT Kanpur.

\begin{figure}[h!]
    \includegraphics[width=0.3\textwidth]{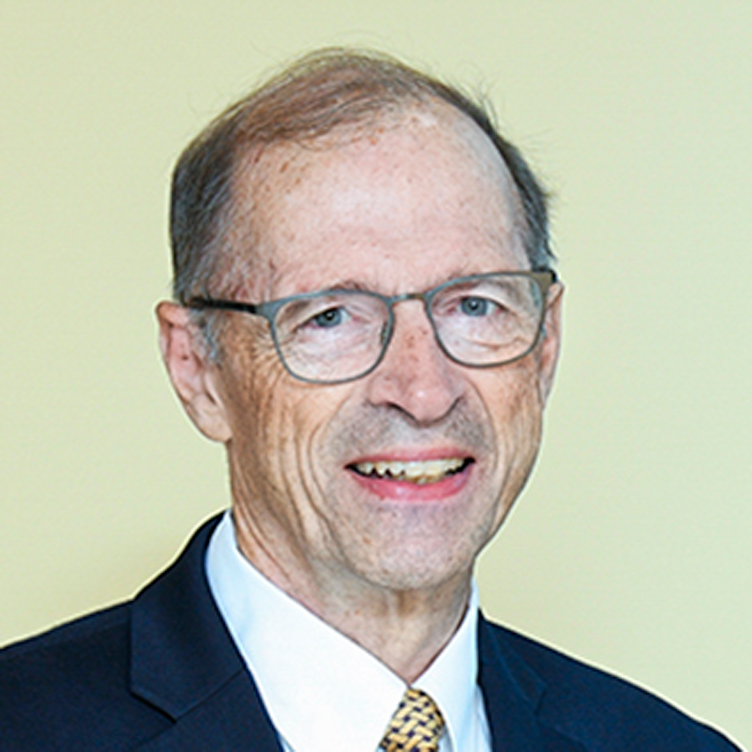}
\end{figure}

\noindent \textbf{George C. Schatz} is the Charles E. and Emma H. Morrison Professor of Chemistry at Northwestern University. He earned his B.S. in Chemistry from Clarkson University and his Ph.D. from Caltech, followed by postdoctoral research at MIT. A faculty member at Northwestern since 1976, Schatz is a theoretical chemist whose work focuses on the optical, structural, and thermal properties of nanomaterials, such as plasmonic nanoparticles, metamaterials, soft matter nanostructures, and transition metal dichalcogenides. His research also encompasses theories of dynamical processes, including gas-phase and surface reactions, electron and energy transfer, quantum phenomena, transport processes, and photochemistry. He has authored over a thousand scientific papers and four books. He is a member of both the National Academy of Sciences and the American Academy of Arts and Sciences, and is a Fellow of the American Physical Society, the Royal Society of Chemistry, the American Chemical Society, and the AAAS. His honors include the Debye, Langmuir, and Marsha Lester Awards from the ACS; the Bourke and Boys–Rahman Awards from the Royal Society of Chemistry; and the Materials Theory Award from the Materials Research Society.

\section{Acknowledgements}
N.S.C. is grateful to Dr. Charles Jason Zeman IV for mentorship and Dr. Yanze Wu for helpful discussions on plasmon decay processes.

\section{Funding}
N.S.C., S.C., A.G., S.K.G., and G.C.S. acknowledge support by NSF grant no. CHE-2347622 for theory development and to the Office of Basic Energy Science, Department of Energy, through grant DE-SC0004752 for theory applications. S.C. acknowledges support from the International Institute for Nanotechnology. N.S.C. gratefully acknowledges support from the National Science Foundation Graduate Research Fellowship through grant no. DGE-22334667. Any opinions, findings, and conclusions or recommendations expressed in this material are those of the authors and do not necessarily reflect the views of the National Science Foundation. Computational time was provided in part by the Quest High-Performance Computing facility at Northwestern University, which is jointly supported by the Office of the Provost, the Office for Research, and Northwestern University Information Technology.

\begin{suppinfo}
    
\end{suppinfo}
\bibliography{bibliography}

\newpage
\begin{figure*}
    \centering
    \includegraphics[width=0.5\linewidth]{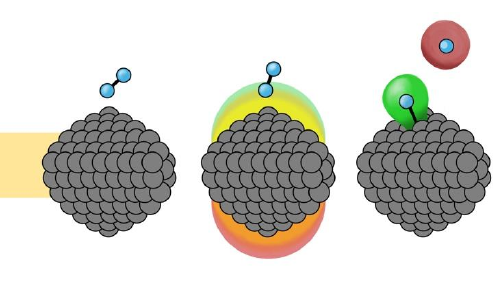}
    \caption{TOC Graphic}
    \nonumber
    \label{fig:enter-label}
\end{figure*}
\end{document}